\documentclass[lettersize,journal]{IEEEtran}

\usepackage{amsmath,amsfonts}
\usepackage{algorithmic}
\usepackage{algorithm}
\usepackage{array}
\usepackage{siunitx}
\usepackage[caption=false,font=normalsize,labelfont=sf,textfont=sf]{subfig}
\usepackage{textcomp}
\usepackage{url}
\usepackage{verbatim}
\usepackage{multirow}
\usepackage[utf8]{inputenc}
\usepackage{graphicx}
\usepackage{epstopdf}
\usepackage{subcaption}
\usepackage{float}
\usepackage{caption} 
\usepackage{authblk}
\usepackage{bm}
\usepackage[font=small]{caption}
\usepackage{xcolor}
\usepackage{hyperref}
\usepackage{amsmath}
\usepackage{makecell}
\usepackage{cite}

\abovedisplayshortskip
\belowdisplayshortskip

\DeclareGraphicsExtensions{.jpg,.png,.pdf}
\begin{document}

\title{QoS Improvement in Multi User Cellular-Symbiotic Radio Network Assisted by Active-STAR-RIS}

\author{Rahman Saadat Yeganeh\,$^1$, Mohammad Javad Omidi\,$^{1,2}$,  
Farshad Zeinali\,$^3$,~\IEEEmembership{Student Member, IEEE},\,\,\,\,\,\,\,\,\,
Mohammad Robat Mili\,$^4$, and Mohammad Ghavami\,$^5$,~\IEEEmembership{Senior Member, IEEE}

\thanks{$^1$Department of Electrical and Computer Engineering, Isfahan University of Technology, Isfahan 84156-83111, Iran (emails: r.saadat@ec.iut.ac.ir, omidi@iut.ac.ir).}
\thanks{$^2$Department of Electronics and Communication Engineering, Kuwait College of Science and Technology, Doha 35003, Kuwait.}
\thanks{$^3$The Pasargad Institute for Advanced Innovative Solutions (PIAIS), Tehran, Iran (email: farshad.zeinali@piais.ir).}
\thanks{$^4$The Pasargad Institute for
Advanced Innovative Solutions (PIAIS), Tehran, Iran (email: Mohammad.Robatmili@gmail.com).}
\thanks{$^5$Electrical and Electronic Engineering Department, London South Bank University, London SE1 0AA, U.K.   (email: ghavamim@lsbu.ac.uk).}}

\maketitle
\begin{abstract}
In this article, we employ active simultaneously transmitting and reflecting reconfigurable intelligent surfaces (ASRIS) to enhance the quality of 6G cellular network services. The network integrates commensal symbiotic radio (CSR) subsystems to facilitate communication between passive Internet of Things (IoT) users and active users, referred to as symbiotic backscatter devices (SBDs) and symbiotic user equipments (SUEs), respectively. Since the SBDs are passive, transmitting information to the SUEs poses significant challenges. To overcome this challenge, we harness the capabilities of massive multiple input multiple output (MIMO) antennas within the base station (BS) to relay the information transmitted by SBDs with greater power. This scheme uses the non-orthogonal multiple access (NOMA) technique for multiple access among all users, and potential interferences are eliminated using successive interference cancellation (SIC).
The primary objective is to maximize the throughput between SBDs and SUEs. To achieve this, we formulate an optimization problem involving variables such as active beamforming coefficients at the BS and ASRIS, phase adjustments of ASRIS, and scheduling parameters between CSR and cellular networks.
To solve this optimization problem, we used three deep reinforcement learning (DRL) methods: proximal policy optimization (PPO), twin delayed deep deterministic policy gradient (TD3), and asynchronous advantage actor critic (A3C). These methods were simulated, and the results demonstrate that A3C, TD3, and PPO have the best convergence speeds and achieve the highest increases in network throughput, respectively. Finally, the proposed scheme was evaluated using passive simultaneously transmitting and reflecting RIS (STAR-RIS), which demonstrated poorer performance compared to ASRIS.

\end{abstract}


\section{Introduction}
\subsection{Background}
The sixth-generation (6G) wireless networks need to be capable of providing coverage for billions of IoT, cellular, and other wireless devices \cite{8879484}. The IoT stands out as a pivotal technology in shaping the future of wireless communications, embracing a diverse range of applications, including smart transportation, smart homes, smart grids, and smart agriculture \cite{you2021towards}. However, the extensive deployment of IoT devices faces two critical challenges: energy efficiency and spectrum scarcity issues \cite{r1,r3,r28}.
The challenge of spectrum scarcity arises from the limited availability of spectrum resources for IoT applications, as a significant portion has already been allocated to various radio systems, including cellular networks, television broadcasts, and other communication services. This constraint underscores the urgency of resolving these challenges to optimize spectrum utilization. Additionally, the energy efficiency issue presents a significant obstacle, given the high cost associated with regularly charging devices or replacing batteries \cite{zhang20196g}.
We propose a solution to address the aforementioned challenges by employing symbiotic radio (SR) systems and reconfigurable intelligent surfaces (RIS) structures. 

SR stands out as an innovative technology that not only preserves the merits of prior systems like cognitive radio\cite{r22,r23,r24} and ambient backscatter communication (AmBC)\cite{r19,r26} but also addresses and eliminates their drawbacks. It currently ranks among the captivating subjects in both scientific and industrial domains \cite{yeganeh2024sum,r20,r21}. 
The SR network can be classified into two types based on the relationship between the symbol periods of the SBDs and the BS: parasitic SR (PSR) and commensal SR (CSR) \cite{8907447,r26}. In PSR, SBDs can exchange information at a high rate, but they also suffers from interference between the signals of SBDs and BS in the receiver, necessitating complex interference cancellation techniques. Furthermore, PSR requires synchronization between the BS, SBDs, and receiver. On the other hand, CSR is suitable for IoT networks with low data rates, and addresses the drawbacks of the PSR setup \cite{r25}. By reducing interference between different network components, the receiver can perform joint decoding of information from both the BS and SBDs, enabled by transmit collaboration between them \cite{r22}.

On the other hand, RIS is a new technology composed of two-dimensional surface that can reduce the need for expensive BS active antennas with complex hardware in the network and provide complete coverage over a specific area. The initial type of these surfaces only had the capability to reflect signals in one direction \cite{wu2021intelligent}, and due to inadequate coverage, a more advanced type called STAR-RIS was introduced in subsequent designs \cite{liu2021star,xu2021star}. STAR-RISs are equipped with a large number of low-cost passive elements whose transmission and reflection coefficients can be controlled.
It is worth emphasizing that utilizing STAR-RIS in passive mode necessitates a substantial number of elements on its surface to attain optimal operational gains, resulting in its enlargement. This presents a formidable challenge to the expansion of these systems. As a result, recent research endeavors have introduced the concept of ASRIS to skillfully address and resolve this issue \cite{xu2023active}.  
\subsection{Related Works}

\subsubsection{Reconfigurable Intelligent Surfaces}
In \cite{niu2021weighted}, a STAR-RIS assisted MIMO system, in which the sum rate maximization problem was investigated in both unicasting and broadcasting signal models. Also, in \cite{li2022enhancing}, an investigation into the secrecy performance of STAR-RIS assisted NOMA networks took place. In \cite{long2021active}, the proposition of active RISs and the investigation of joint optimization for reflecting phase shift and receive beamforming were introduced. Subsequently, \cite{zhang2022active} presented a validated signal model for active RISs. In the article \cite{xu2023active}, a hardware model is introduced for active STAR-RISs incorporating both coupled and independent transmission and reflection phase-shifts. More precisely, this paper demonstrates the utilization of reflection-type amplifiers and quadrature hybrid couplers to achieve amplification of the transmission and reflection coefficients. Nevertheless, a significant gap exists in the research on modeling and performance analysis of ASRISs. Furthermore, the strategy for utilizing ASRIS to improve Quality of Service (QoS) in communications within an operational IoT and cellular network is currently ambiguous.

\subsubsection{Symbiotic Radio}
The paper \cite{gu2024breaking} provides a comprehensive review of backscatter communication, addressing key challenges such as interference and double-path fading, while exploring advancements and future trends to enhance its role in the evolving IoT ecosystem. In \cite{yeganeh2023multi}, the advanced version of backscatter communication, known as symbiotic radio, has been implemented in the CSR form. This system is specifically designed for scenarios involving multiple backscatter devices (BDs). The primary goal within this framework is to reduce energy consumption, and a method for optimal resource allocation, named Timing SR, has also been introduced. Additionally, the paper \cite{wang2024multi} presents a system model similar to the one proposed in \cite{yeganeh2023multi}, considering a multi-user scenario at the destination. The aim of this paper is to design transmit beamforming that minimizes the transmit power at the BS while adhering to a cellular transmission outage probability constraint.
The study in \cite{yang2023novel} introduces a symbiotic backscatter NOMA (SBN) system, where a backscatter device transmits information to two users by adjusting its reflection coefficient, ensuring successful decoding of both NOMA and backscatter signals. Similarly, \cite{yang2023ris} proposes a segmented RIS-based symbiotic ambient backscatter system that improves signal reception and backscatter performance, optimizing the coexistence outage probability and ergodic capacity. Building on these concepts, the work in \cite{liu2024star} presents a STAR-RIS segmented SBN system, which optimizes reflection coefficients to maximize ergodic capacity while balancing the decoding performance of primary and backscatter signals. Further, \cite{wen2024outage} investigates the outage performance of the STAR-RIS segmented SBN system, analyzing the effects of interference and power allocation. Additionally, the letter \cite{yang2022opportunistic} proposes two mechanisms for backscatter devices to enhance concurrent transmissions in primary and backscatter systems.

Also, the article \cite{8907447}, proposes a novel SR technique for passive IoT devices. This approach involves integrating a BD with a primary communication system, and designing a primary transmitter and receiver to optimize both the primary and BD transmissions. The decoding strategy used in the receiver is based on SIC.
So far, articles on SR have primarily focused on IoT device communications. However, the main goal of the SR system is to integrate it into a cellular network. This integration presents various challenges and requires further examination. Furthermore, to efficiently accommodate a significant number of devices in SR networks, it is crucial to design the network to support multiple SBDs. However, challenges, including potential user interference with improperly designed multiple access schemes, may arise. To mitigate these issues, various methods, such as NOMA, are employed to prevent interference \cite{wu2023ris,li2023physical,yang2023novel}.

It should be noted that, SR systems typically suffer from the poor system performance owing to the limited communication efficiency of backscattering devices. To overcome this bottleneck, one promising method involves using RIS to enhance the received signal strength in SR transmission. Due to the abundant advantages of these two subjects and their combination, numerous articles, such as \cite{hu2022reconfigurable,hu2020reconfigurable,ye2021capacity}, have been presented in this field.
Also, in the article \cite{zhou2023transmit}, a passive STAR-RIS empowered transmission scheme for SR systems. The authors focused on minimizing the transmit power of the BS by designing the active beamforming and simultaneous reflection and transmission coefficients under the practical phase correlation constraint.

Another method involves utilizing the power of a active antennas on BS to relay information from IoT devices. In this scheme, the BS receives data from IoT devices and forwards it to the intended destination. This method has been demonstrated for a single user in \cite{gong2018backscatter}. In this paper, to enhance the BS signals for transmission to the destination, a backscatter relay scheme is employed, where multiple passive devices relay the BS information. This approach introduces challenges due to increased interference at the destination and the inherent limitations of passive devices in signal amplification and transmission, particularly in double fading channels.

\subsection{Our Contributions}
In this work, we propose a 6G cellular network featuring a CSR subsystem. The BS initially transmits signals to the SBDs on one side using active massive MIMO antennas. The SBDs modulate their information onto the received signal carrier. Due to the significant distance between the passive IoT devices and their intended recipients (SUEs), we employ a relay system. The BS relays the information to the SUEs located on the opposite side of the BS.
To enhance the performance of user data transmission in this network, we incorporate an ASRIS device.  This device not only amplifies the received signals but also transmits them to users within its coverage area. These users typically lack a line-of-sight (LoS) connection with the BS due to intervening obstacles. In all transmission schemes between nodes, the method for multiple access is NOMA.

 The major contributions of this paper are summarized as follows:
\begin{itemize}
\item{First, We implement an innovative, comprehensive, and practical system model where multiple users, both actively (SUEs) and passively (SBDs), engage in the exchange of diverse information within the 6G cellular network. To facilitate communication between IoT users and other cellular devices, we employ a CSR setup. In this setup, the BS dispatches ambient signals to SBDs using a NOMA approach. The SBDs then harvest wireless energy from these signals, modulate their own information onto the carrier, and transmit backscatter signals back to the BS. Assisted by an ASRIS, the BS efficiently relays this information to the intended SUEs located in both the reflection and transmission regions of the ASRIS, ensuring effective communication for users with LoS and non-LoS links to the BS, respectively.}

\item{Second, To achieve optimal resource allocation, we formulate an optimization problem aimed at maximizing the information throughput among SBDs, the BS, and SUEs. To this end, we define several key variables within the problem, including the active beamforming vectors for two phases: transmitting signals to the SBDs and delivering the desired information to the SUEs. We also consider a timing schedule for these phases and determine the amplifier gain and phase variables for the ASRIS. To ensure QoS, we impose constraints to guarantee a minimum rate for each SBD and SUE. Additionally, we incorporate constraints related to SIC for all transmitted signals from the users.
}
\item{Third, SBDs can harvest the required energy from ambient waves whenever they need to transmit information. Due to the considerable distance between SBDs and SUEs, direct transmission to SUEs may not be feasible. To address this issue, we leverage the power of active massive MIMO antennas at the BS. After eliminating interference using SIC, the BS directs each of the signals from the SBDs to their respective SUEs. Additionally, the signal modeling in ASRIS is thoroughly examined and incorporated into the overall design.
}

\item{Finally, to solve this complex problem, we employ DRL methods. Three novel approaches, namely PPO, TD3, and A3C, each exhibiting distinctive features, were explored. These algorithms are employed to model the problem, and a comprehensive comparison of all simulated methods is conducted under varying conditions.}
\end{itemize}

This paper is structured as follows. In Section II, we present the proposed system model for the ASRIS assisted by the CSR system. Section III focuses on the throghput maximization problem. In Section IV, we investigate some DRL methods, namely PPO, TD3 and A3C. In Section V, the mentioned methods are modeled and simulated, followed by a comparison with each other. Finally, in Section VI, we summarize our conclusions and discuss future work.

\textbf{\emph{Notations}}: ${{\bf{A}}^H}$, ${{\bf{A}}^T}$, $\left\| {\bf{A}} \right\|$ 
  denote the trace, conjugate transpose, transpose, and norm of the matrix ${\bf{A}}$, respectively. Also, the gradient operator denoted as $\nabla$.

\section{SYSTEM MODEL AND PROBLEM FORMULATION}
\begin{figure}
\includegraphics[width=8.7cm]{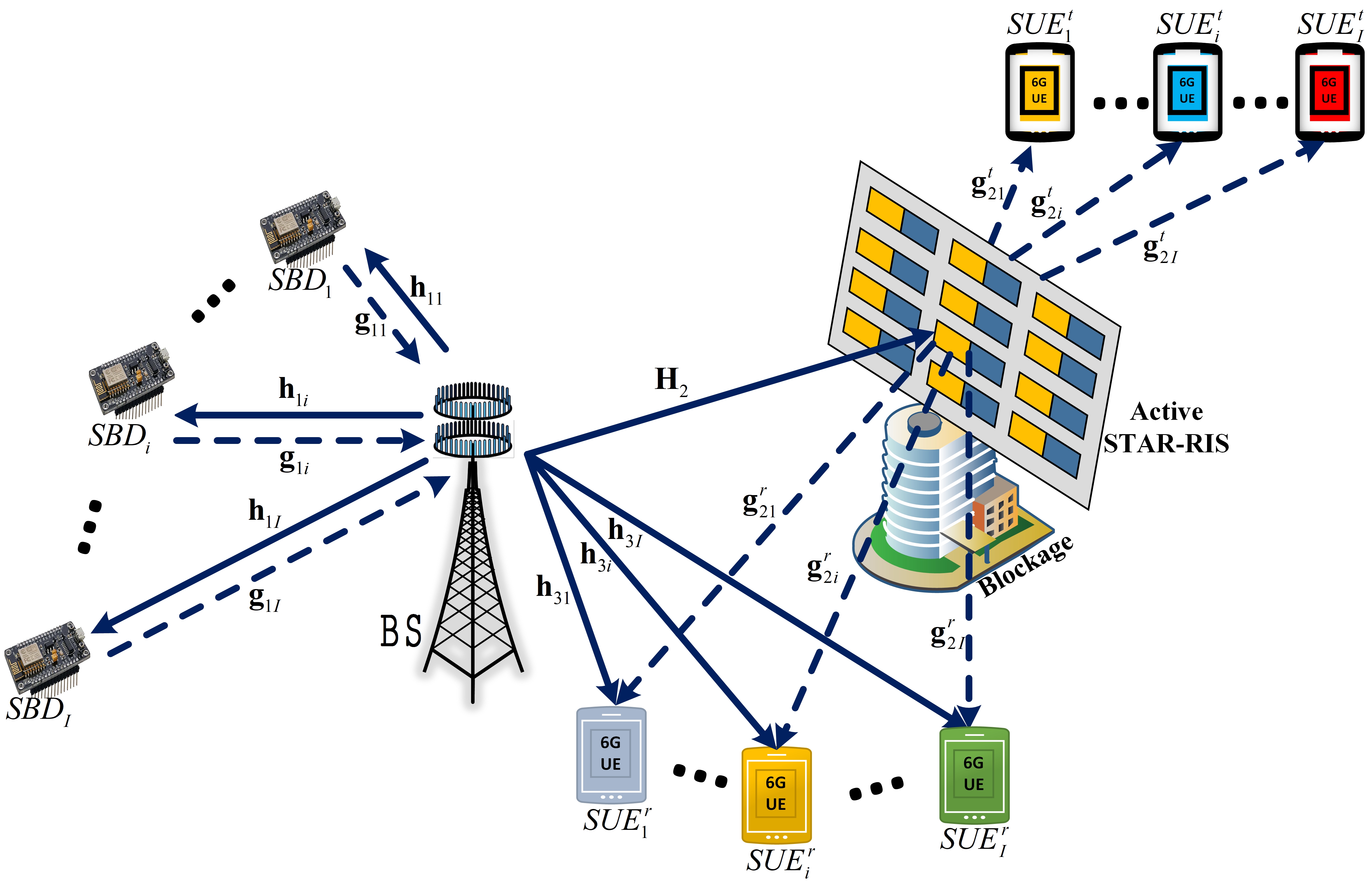}
\caption{Symbiotic radio system model with SBDs, STAR-RIS and SUEs.} 
\label{fig:STAR_RIS_scheme 4}
\end{figure}

As illustrated in the system model in Fig. \ref{fig:STAR_RIS_scheme 4}, the cellular network with the SR subsystem comprises a BS equipped with ${N}$ massive MIMO antennas, an ASRIS with ${M}$ active elements, $I$ SUEs in the transmission and reflection spaces of the ASRIS, and $I$ SBDs with a single antenna.

In this structure, the SBDs are positioned at a long distance from the SUEs (relative to the wavelength of the network's operating frequency). Moreover, each SBD is a passive device and lacks the capability to transmit information over long distances, it cannot establish a direct backscatter communication link with the SUEs.
Therefore, to establish communication, we need to use the cooperative relay structure. An appropriate idea for this is to use the BS capabilities, which is located in the center of the cell and can receive and amplify the SBD's signal. This amplified signal reaches the SUEs through a direct link and with the help of ASRIS. As a result, the communication between SBDs and SUEs is established through the cooperation of BS and ASRIS. This method significantly increases the range and quality of communication.

In this section, we analyze the signal model for communication between the SBDs, BS, ASRIS, and SUEs. The BS establishes a direct communication link with $\text{SUE}_j$, $ \text{j}\in\{1, 2, \ldots, i, \ldots, I\}$ in reflection space of ASRIS, while there is no direct link between BS and the $\text{SUE}_j$
 in transmission space of ASRIS.

Since the primary objective of this paper is to explore the implementation of IoT users communications leveraging the infrastructure and spectrum of a cellular network, and to examine how components of this infrastructure, including the BS and ASRIS, can enhance the reliability and stability of SBD and SUE communications, we have chosen to exclude the study of intra-network communications within the cellular network itself (i.e., communications between the BS and SUEs). It is also worth noting that, given the minimal data volume generated by SBD users, they are unlikely to have any adverse impact on the cellular network under practical conditions.

\begin{table}
	\centering
	    \caption{List of abbreviations.}
	\begin{tabular}{|c|c|}

				\hline  AmBC & Ambient backscatter communication    \\ 
				\hline    ASRIS & Active simultaneously transmitting and reflecting RIS    \\ 
				\hline  A3C & Asynchronous advantage actor critic\\ 
				\hline  BS  & Base station \\ 
				\hline  CSR & Commensal symbiotic radio\\ 
				\hline DDPG & Deep deterministic policy gradient \\
				\hline  DRL & Deep reinforcement learning\\ 
				\hline  IoT & Internet of things\\ 
				\hline  MIMO & Multiple input multiple output\\ 
		\hline  NOMA & Non orthogonal multiple access   \\
		\hline  PPO & Proximal policy optimization\\ 
		\hline  QoS & Quality of Service\\ 
		\hline  RIS & Reconfigurable intelligent surfaces\\
		\hline  SIC & Successive interference cancellation\\ 
		\hline  STAR-RIS & Simultaneously transmitting and reflecting RIS\\ 
		\hline  SR & Symbiotic radio\\  
		\hline  SBD & Symbiotic backscatter device\\ 
		\hline  SUE & Symbiotic user equipment\\ 
		\hline  TD3 & Twin delayed DDPG\\ 

		\hline 

	\end{tabular} 
\end{table}

\subsection{System Model of the Active STAR-RIS}

The STAR-RISs are advanced devices that enable independent control of the transmitted and reflected signals. Specifically, the signal incident upon the $m$th element of the ASRIS is denoted by $s_m$, where $m \in \mathcal{M} \buildrel \Delta \over = \{ 1,2,...,M\}$. The $m$th element can adjust the amplitude and phase of the incident signal during transmission and reflection, resulting in transmitted and reflected signals given by $\left(\sqrt{\beta_m^t}e^{j\theta_m^t}\right)s_m$ and $\left(\sqrt{\beta_m^r}e^{j\theta_m^r}\right)s_m$, respectively. Here, $\beta_m^t$ and $\theta_m^t$ represent the amplitude and phase shift adjustments made by the $m$th element during transmission, while $\beta_m^r$ and $\theta_m^r$ represent the corresponding adjustments during reflection. The signals transmitted and reflected by each element of the ASRIS can be accurately modeled with this approach \cite{zuo2022joint}:
\begin{equation}
\label{eq:t_m}
{t_m} = \left( {\sqrt {\beta _m^t} {e^{j\theta _m^t}}} \right){s_m}, \quad \forall m \in \mathcal{M}
\end{equation}
\begin{equation}
\label{eq:t_r}
{r_m} = \left( {\sqrt {\beta _m^r} {e^{j\theta _m^r}}} \right){s_m}, \quad \forall m \in \mathcal{M}.
\end{equation}

%\ref{eq:t_r}

In the ASRIS structure each element can operates in simultaneous transmission and reflection mode, which is more general than either full transmission mode or full reflection mode \cite{liu2021star, mu2021simultaneously}. As a result, we adopt the operating protocol for the ASRIS known as energy splitting, where all elements of the ASRIS are assumed to operate in transmission and reflection mode. The energy splitting mode provides more degrees of freedom for optimizing the network, but it also increases the communication overhead between the BS and ASRIS \cite{mu2021simultaneously}. In order to reduce complexity and provide greater clarity, we propose a simplified equal energy-splitting protocol, similar to that in \cite{10013760}. This protocol involves setting \(\beta _m^t = \beta _m^r = \frac{{{p_{\text{ASRIS}}}}}{2}\) in both transmission and reflection modes, where ${p_{\text{ASRIS}}}$ is the power ASRIS amplifier, intended for the elements, remains constant over time.

For ease of expression, we define the transmission (when $l=t$) or reflection (when $l=r$) beamforming vector as $\mathbf{R}_l = [\sqrt{\beta_1^l}e^{j\theta_1^l}, \sqrt{\beta_2^l}e^{j\theta_2^l}, \ldots, \sqrt{\beta_M^l}e^{j\theta_M^l}]^H\in \mathbb{C}^{M\times 1}$, and let $\boldsymbol{\Theta}_l = \mathrm{diag}(\mathbf{R}_l^H)\in \mathbb{C}^{M\times M}$ be the corresponding diagonal beamforming matrix of ASRIS.

In the context of the passive structure of STAR-RIS, it is important to note that according to the energy conservation law, the energy of the incident signal must equal the sum of the energies of the transmitted and reflected signals, expressed as $|s_m|^2 = |t_m|^2 + |r_m|^2$ for all $m \in \mathcal{M}$. This implies a coupling between the amplitudes of the transmission and reflection coefficients, necessitating that $\beta_{m}^{t} + \beta_{m}^{r} = 1$. 

\subsection{Problem Formulation}
We propose the implementation of a CSR as the considered structure for the SR network. In this configuration, the BS transmits $K$ symbols (where $k=1,2,...,K$ and $K\gg 1$) for every SBD's data symbol. This means that the duration of each symbol transmitted by SBD ($T_{\text{SBD}}$) is equal to $K$ times of the duration of symbol transmission by BS ($T_{\text{BS}}$). With this CSR design, we ensures that the SBDs and BS signals do not interfere with each other at the destination.

\begin{figure}
\includegraphics[width=8.7cm]{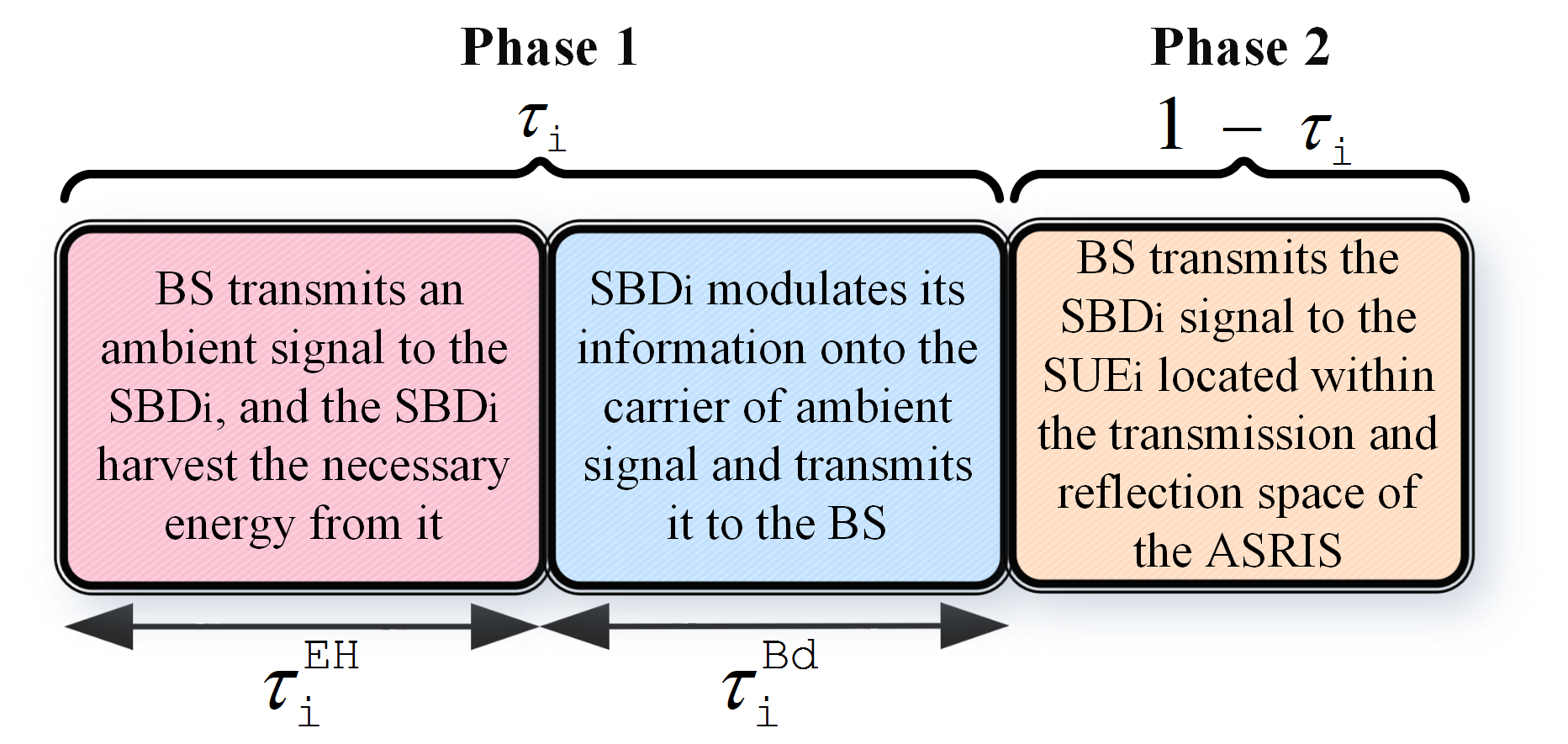}
\caption{The time division duplexing in the 6G Cellular-CSR network.} 
\label{fig:TDD_frame}
\end{figure}

In the proposed cellular-CSR network, time division duplexing for sending and receiving information from the BS to the SBDs and then from SBDs to BS and SUEs is considered in two phases, which occurs within a time unit of 1. This is illustrated in Fig. \ref{fig:TDD_frame}. In the first phase, the CSR network is implemented, and the information from the SBDs reaches the BS. In the second phase, this information is sequentially delivered to its intended destination within a cellular network. The signal analysis for each phase is as follows.

In the first phase, the BS transmits the known signal $s_{k}(t)$ to $\text{SBD}_i$ through the complex channel ${\bf{h}}_{\text{1i}}^{H} \in {{\Bbb C}^{1 \times N}}$. This transmission is achieved using the transmit beamforming vector ${{\bf{w}}_{\text{1i}}} \in \mathbb{C}^{N\times 1}$ and with power $P_i$. 
The received signal at $\text{SBD}_i$ can be obtained as follows:
\begin{equation}
\label{eq:y_SBD}
y_k^{{\text{SBD}_{i}}}\left( t \right) = \sqrt {{P_i}} {\bf{h}}_{\text{1i}}^H{{\bf{w}}_{\text{1i}}}{s_k}\left( t \right) + n_k^{{\text{SBD}_{i}}}\left( t \right),\forall i \in \psi,
\end{equation}
where \(\psi\buildrel \Delta \over = \{1,2,...i,...,I\}\) represents the collection of all SBDs or SUEs. Upon receiving the ambient signal $y_k^{SB{D_i}}\left( t \right)$, the SBDs harvests energy for their electronic circuits. These steps are carried out within the time duration $ {\tau _i^{\text{EH}}}$. 

In this paper, we utilize NOMA to facilitate multiple access for SBDs and SUEs, optimize resource allocation among them, and mitigate excessive interference at the destination.  Therefore, the BS transmits signals to each of the SBDs with different power levels based on their distances and channels gain. As shown in Fig. \ref{fig:STAR_RIS_scheme 4}, the allocated power to the SBDs is in the form of \({P_I} > ... > {P_i} > ... > {P_1}\). Here, we assume that the transmitted symbol from the ambient signal sent to all SBDs is the same and equal to $s_{k}(t)$.

After the SBDs are charged with the ambient signal $y_k^{\text{SBD}_{i}}\left( t \right)$, they subsequently modulate their information, denoted as $c_i(t)$, $i \in \psi$, onto the ambient carrier signal. It then backscatters the modulated signal back to the BS using the SBD's reflection coefficient ${\eta_i}$, \(0 \le {\eta _i} \le 1\). These steps are carried out within the time duration $ {\tau _i^{\text{Bd}}}$. It should be noted that since SBDs does not have any power-consuming active components, $n_k^{\text{SBD}_i}\left( t \right)$, which is the complex gaussian noise at the SBD's antenna, is very small and can be disregarded \cite{r41}. 
Assuming that the beamforming vector used during signal reception at the BS is denoted as $ {{\bf{w}}_{Ri}^H} \in \mathbb{C}^{1 \times N}$, the received signal at the BS can be expressed as follows:
\begin{equation}
\label{eq:y_BS}
y_k^{\text{BS}}\left( t \right) = {s_k}\left( t \right)\sum\limits_{j \in \psi } {\sqrt {{P_j}{\eta _j}} {\bf{w}}_{Rj}^H{{\bf{g}}_{\text{1j}}}{\bf{h}}_{\text{1j}}^H{{\bf{w}}_{\text{1j}}}{c_j}\left( t \right)}  + n_k^{\text{BS}}\left( t \right),
\end{equation}
where \(n_k^{BS} \sim \mathcal{CN}\left( {0,\sigma _{BS}^2} \right)\) is the circularly symmetric complex gaussian (CSCG) noise at the BS and ${{\bf{g}}_{\text{1i}}}\in \mathbb{C}^{N\times 1}$ is the complex channel vector from $\text{SBD}_i$ to BS.  

In this paper, we assume that all channels experience are flat fading and remain constant within a given time frames. Channel estimation for active users is typically performed using the pilot signal method. For passive users in SR networks, several channel estimation techniques can be applied, some of which are explored in the papers \cite{RN120}, \cite{RN119}, \cite{RN121}. Therefore, because it is beyond the scope of this paper, we assume that the channel state information (CSI) for all channels is readily available.

The BS is designed with a robust infrastructure and powerful processors, which enable it to perform complex signal processing tasks. Moreover, since the signal $s_{k}(t)$ is specific to the BS and well known, and given the availability of CSI for channels ${\bf{g}}_{\text{1i}}$ and ${\bf{h}}_{\text{1i}}^{H}$, the BS is capable of efficiently extracting the information $c_{i}(t)$ for $\text{SBD}_i$ in its baseband and remove any other noise from it. Furthermore, to eliminate interference caused by other SBDs, which have a stronger signal to interference plus noise ratio (SINR) compared to $\text{SBD}_i$ and simultaneously transmit their signals to the BS, the SIC method is employed.

So far, the desired signal for $\text{SBD}_i$ has been successfully received and decoded by the BS.  To enhance the signal strength and improve reception quality, the maximal ratio combining (MRC) technique is applied at the BS receiver. The corresponding beamforming vector is given by 
\(
\mathbf{w}_{Ri} = \frac{\mathbf{g}_{\text{1i}}}{\| \mathbf{g}_{\text{1i}} \|}
\).

In the CSR setup, a single SBD symbol is transmitted over \( K \) consecutive BS symbol intervals, enabling the primary signal \( s_k{(t)} \) to be interpreted as a spread-spectrum code of length \( K \) for the SBD symbols. Consequently, the SINR for decoding the SBD symbol \( c_i{(t)} \) increases by a factor of \( K \), though this comes at the cost of reducing the symbol rate by a factor of \( 1/K \). Therefore, by considering Eq. \ref{eq:y_BS} and assuming \( \mathbb{E}\left[ \left| {s_{k}(t)} \right|^2 \right] = 1 \), the achievable rate, denoted as \( R_{{\rm{SB}}{{\rm{D}}_i}}^{1} \), after applying the SIC technique at the BS during the first phase, can be determined as follows:

\begin{equation}
\label{eq:y_R_1}
R_{{\rm{SB}}{{\rm{D}}_i}}^1 = \frac{{B\tau _i^{Bd}}}{K}{\log _2}\left( {1 + \frac{{K{P_i}{\eta _i}{{\left| {{{\bf{g}}_{{\rm{1i}}}}{\bf{h}}_{{\rm{1i}}}^H{{\bf{w}}_{{\rm{1i}}}}} \right|}^2}}}{{\sum\limits_{j \in \upsilon } {{P_j}{\eta _j}{{\left| {{{\bf{g}}_{{\rm{1j}}}}{\bf{h}}_{{\rm{1j}}}^H{{\bf{w}}_{{\rm{1j}}}}} \right|}^2}}  + B\sigma _{{\rm{BS}}}^2}}} \right),
\end{equation}
where $B$ represents the bandwidth of the receiver filter in the BS. Without loss of generality, for the sake of simplicity, we can consider the bandwidth is $B=1$ Hz in all scenarios. Furthermore, ${\tau _i} = \tau _i^{\text{EH}} + \tau _i^{\text{Bd}}$ denotes the allocated time for completing the first phase. Also,
 \(\upsilon  \buildrel \Delta \over = \{ 1,...,i - 1\} \) represents the set of all SBDs whose distance to the BS is less than that of $\text{SBD}_i$, and consequently, these users receive lower power allocation. These SBDs cause interference on $\text{SBD}_i$, whose total interference is shown in the denominator of the SINR fraction. 
It is worth noting that, since in this design we aim to examine the data exchange rate of users, we conduct this analysis in the worst-case scenario, which involves the simultaneous reception of signals from all SBDs at the BS. It is obvious that in a real and operational scenario, where signals are not received simultaneously, there will be less interference, and as a result, the output of the design will be more favorable.

As stated, SBDs must first harvest the necessary amount of energy wirelessly from ambient waves in order to establish communication. According to Eq. \ref{eq:y_SBD}, the energy that can be harvested by ${\text{SB}}{{\text{D}}_{\text{i}}}$ (${\varepsilon _{\text{SBD}_{i}}}$) during the time slot \(\tau _i^{\text{EH}}\), can be expressed as follow:
\begin{equation}
 {\varepsilon _{\text{SBD}_{i}}} \le {\Gamma _i} {P_i}{\tau _i^{\text{EH}}}(1 - {\eta _i})|{\bf{h}}_{\text{1i}}^H{{\bf{w}}_{\text{1i}}}{|^2}, \,\,\,\,\,\,\,\, i \in \psi
\end{equation}where, \(0 \le {\Gamma_i } \le 1\)  is the  energy conversion efficiency by $\text{SBD}_{i}$. The maximum energy harvested by ${\text{SB}}{{\text{D}}_{\text{i}}}$, occurs when the power reflection coefficient $\eta _i$ is equal to zero.

Although ASRIS could indeed be used to support and improve SBD communications, we refrained from using it in this phase to reduce the complexity of the mathematical relationships. Additionally, since we are investigating the effects of using ASRIS in the cellular network (second phase) and the same effects can occur in the assumed CSR network, we chose not to apply ASRIS here.

In the second phase, the BS  transmits the signal $c_{i}(t)$, $i \in \psi$ to the destinations ASRIS and $\text{SUE}_i$, $i \in \psi$. Specifically, $\text{SUE}_i$ is located in the reflection space of ASRIS, denoted as $\text{SUE}_{i}^{r}$. Both ASRIS and $\text{SUE}_{i}^{r}$ have direct links with the BS. The transmission is carried out with a power $P_i$ to $\text{SUE}_{i}^{r}$ using the beamforming vector $\mathbf{w}_{\text{2i}}\in \mathbb{C}^{N\times 1}$. In this section, we also utilize NOMA for accessing multiple SUEs present in the reflecting space of ASRIS. In this scenario, SUEs are located at different distances from BS and ASRIS, and therefore, the power of the transmitted signal is allocated based on their distance and channels gain. In this scenario, the received signal that reaches the ASRIS through the channel 
${\mathbf{H}_2^H} \in {{\Bbb C}^{M \times N}}$ is given by:
\begin{equation}
\label{eq:y_ASRIS}
{{\bf{y}}_{\text{ASRIS}}}\left( t \right) = {\mathbf{H}_2^H}\sum\limits_{j\in \psi } {\sqrt {{P_j}} {{\bf{w}}_{\text{2j}}}{c_j}\left( t \right)}  + {{\bf{n}}_{\text{ASRIS}}}(t).
\end{equation}

As mentioned above, since the ASRIS has active components, the presence of CSCG noise, denoted as ${{\bf{n}}_{{{\text{ASRIS}}}}} \sim \mathcal{CN}\left( {{\bf{0}},\sigma _{\text{ASRIS}}^2{{\bf{I}}_{\bf{M}}}} \right)$, should be taken into account.

On the other hand, considering Fig. \ref{fig:STAR_RIS_scheme 4}, the received signal at $\text{SUE}_{i}^{r}$, which arrives through both the direct link from the BS and the reflected signal from ASRIS, can be obtained as follows:
\begin{equation}
\label{eq:y_SUE1}
\begin{aligned}
y_{\text{SUE}_i}^r(t) &= 
\left( \mathbf{g}_{2i}^r \mathbf{\Theta}_r \mathbf{H}_2^H + \mathbf{h}_{3i}^H \right)
\sum_{j \in \psi} \sqrt{P_j} \mathbf{w}_{2j} c_j(t) \\
&\quad + \mathbf{g}_{2i}^r \mathbf{\Theta}_r \mathbf{n}_{\text{ASRIS}}(t) + n_{\text{SUE}_i}^r(t),
\end{aligned}
\end{equation}

where ${\bf{g}}_{\text{2i}}^{r} \in {{\Bbb C}^{1 \times M}}$ is the channel vector between BS and $\text{SUE}_{i}^{r}$ in the reflection space of the ASRIS, ${\bf{h}}_{\text{3i}}^{H} \in {{\Bbb C}^{1 \times N}}$ is the channel vector for the direct link between BS and $\text{SUE}_{i}^{r}$, and \({ n_{{\text{SUE}_{i}}}^r}\sim \mathcal{CN}\left( {0,\sigma _{{{{\text{SUE}_{i}^{r}}}}}^2} \right)\) is the CSCG noise at the $\text{SUE}_{i}^{r}$.

According to (\ref{eq:y_SUE1}) the maximum achievable rate after using the SIC technique in the $\text{SUE}_{i}^{r}$, denoted as $R_{{\rm{SU}}{{\rm{E}}_i}}^{2,r}$ is
\begin{equation}
\label{eq:R_2 - s1r}
R_{{\text{SU}}{{\text{E}}_i}}^{2,r} = \left( {1 - {\tau _i}} \right){\log _2}\left( {1 + \gamma _{{\rm{SU}}{{\rm{E}}_i}}^{2,r}} \right),
\end{equation}where
\begin{equation}
\gamma _{{\rm{SU}}{{\rm{E}}_i}}^{2,r} = 
\frac{
P_i {\bf{w}}_{2i}^2 \left| \overline{{\bf{h}}_i^r} \right|^2
}{
\left| \overline{{\bf{h}}_i^r} \right|^2 \sum\limits_{j \in \upsilon} P_j {\bf{w}}_{2j}^2 
+ {\left\| {{\bf{g}}_{2i}^r{{\bf{\Theta }}_r}} \right\|^2}\sigma _{{\rm{ASRIS}}}^2 + \sigma _{{\rm{SUE}}_i^r}^2
},
\end{equation}
{where $1-{\tau _i} $ denotes the allocated time for completing the second phase and \(\overline {{\bf{h}}_{{i}}^{{r}}}  \buildrel \Delta \over = {\bf{g}}_{{\rm{2i}}}^r{{\bf{\Theta }}_r}{\bf{H}}_2^H + {\bf{h}}_{{\rm{3i}}}^H\). In this relation, \(\upsilon\) represents the set of all SUEs in reflection space, whose channels gain (BS-$\text{SUE}_{i}^{r}$ and  BS-ASRIS-$\text{SUE}_{i}^{r}$) is higher than $\text{SUE}_{i}^{r}$ and as a result have lower power allocation.

Also, due to obstacle, a direct link between the BS and $\text{SUE}_{i}$ in the transmission space ($\text{SUE}_{i}^{t}$) of ASRIS is not available.
Consequently, only ASRIS has the capability to transmit the signal to $\text{SUE}_{i}^{t}$ using its own transmission space. It's important to note that if any other structure, such as an active or passive RIS, were used instead of ASRIS,  $\text{SUE}_{i}^{t}$ would be located in a cellular blind spot. In this scenario, the received signal at  $\text{SUE}_{i}^{t}$ , which arrives through the transmission link by ASRIS, can be obtained as follows:
\begin{equation}
\label{eq:y_SUE2}
\begin{array}{*{20}{l}}
{y_{{\rm{SU}}{{\rm{E}}_i}}^t\left( t \right) = {\bf{g}}_{{\rm{2i}}}^t{{\bf{\Theta }}_t}{\bf{H}}_2^H\sum\limits_{j \in \psi } {\sqrt {{P_j}} {{\bf{w}}_{{\rm{2j}}}}{c_j}\left( t \right)}}\\
{{\kern 1pt} \,\,\,\,\,\,\,\,\,\,\,\,{\kern 1pt} {\kern 1pt} {\kern 1pt} {\kern 1pt} {\kern 1pt} {\kern 1pt} {\kern 1pt} {\kern 1pt} {\kern 1pt} {\kern 1pt} {\kern 1pt} {\kern 1pt} {\kern 1pt} {\kern 1pt} {\kern 1pt} {\kern 1pt} {\kern 1pt} {\kern 1pt} {\kern 1pt} {\kern 1pt} {\kern 1pt} {\kern 1pt} {\kern 1pt}  + {\bf{g}}_{{\rm{2i}}}^t{{\bf{\Theta }}_t}{{\bf{n}}_{{\rm{ASRIS}}}}\left( t \right) + n_{{\rm{SU}}{{\rm{E}}_i}}^t(t),}
\end{array}
\end{equation}
where ${\bf{g}}_{\text{2i}}^{t} \in {{\Bbb C}^{1 \times M}}$ is the channel vector between BS and $\text{SUE}_{i}^{t}$ in the transmission space of the ASRIS, and \( n_{{\text{SUE}_{i}}}^t\sim \mathcal{CN}\left( {0,\sigma _{{{{\text{SUE}_{i}^{t}}}}}^2} \right)\) is the CSCG noise at the $\text{SUE}_{i}^{t}$.

According to (\ref{eq:y_SUE2}) the maximum achievable rate after using the SIC technique in the $\text{SUE}_{i}^{t}$, denoted as $R_{{\rm{SU}}{{\rm{E}}_i}}^{2,t}$ is
\begin{equation}
\label{eq:R_2 - s1t}
R_{{\text{SU}}{{\text{E}}_i}}^{2,t} = \left( {1 - {\tau _i}} \right){\log _2}\left( {1 + \gamma _{{\rm{SU}}{{\rm{E}}_i}}^{2,t} } \right),
\end{equation}
where
\begin{equation}
\gamma _{{\rm{SU}}{{\rm{E}}_i}}^{2,t} = 
\frac{
{P_i {\bf{w}}_{2i}^2 {\left| \overline{{\bf{h}}_i^t} \right|}^2}
}{
 {\left| \overline{{\bf{h}}_i^t} \right|}^2 \sum\limits_{j \in \upsilon} P_j {\bf{w}}_{2j}^2
+ {\left\| {{\bf{g}}_{2i}^t{{\bf{\Theta }}_t}}\right\|^2}\sigma _{{\rm{ASRIS}}}^2  + \sigma _{{\rm{SUE}}_i^t}^2
},
\end{equation}where \(\upsilon  \buildrel \Delta \over = \{ 1,...,i - 1\} \) represents the set of all SUEs whose distance to ASRIS is less than $\text{SUE}_{i}^{t}$ and as a result have higher channel gain. Also, \(\overline {{\bf{h}}_i^t}  \buildrel \Delta \over = {\bf{g}}_{{\rm{2i}}}^t{{\bf{\Theta }}_t}{\bf{H}}_2^H\).

It should be noted that in this article, all communication channels with ASRIS are considered to be Rician, and all communication channels with BS are considered to be Rayleigh. Additionally, due to the direct LoS between BS and ASRIS, we also consider this channel to be Rician.

Furthermore, the assumption of an equal number of SUEs and SBDs is made purely for clarity and simplicity in formulating the equations. Any variation in these numbers would impact the interference levels for each user but would not alter the overall concept presented in the paper.

\section{THROUGHPUT MAXIMIZATION}
Based on the explanations provided in the preceding section and in accordance with equations \eqref{eq:y_R_1}, \eqref{eq:R_2 - s1r}, and \eqref{eq:R_2 - s1t}, the main objective of this article is to maximize the throughput in the network. This objective is achieved when the minimum information rate between users in Phase 1 and Phase 2 is maximized. To accomplish this, we define the optimization problem as follows:
\begin{subequations}
\label{eq:main_rate}
\begin{align}
&\max\limits_{{\eta _i},{\tau _i},\tau _i^{{\rm{EH}}},\tau _i^{{\rm{Bd}}},{P_i},\beta _m^l,\theta_m^l} \min \left( {{{\rm{R}}_{{ \text{SBD}_i}}^1}{\rm{ }},R_{\text{SUE}_i}^{2,l}} \right) \tag{14}\\
s.t. \nonumber\\
& \left( {|\beta _m^t{|^2},|\beta _m^r{|^2}} \right) \le \frac{{{p_{\text{ASRIS}}}}}{2},\,\,\,\,{p_{\text{ASRIS}}} = {\rm{cte}}\label{eq:subeq11a}\\
& 0 \le \theta _m^l \le 2\pi ,\;\;\;{\kern 1pt}  \,\,\,\,\,\,\,\, \forall m \in {\cal M}\;\;\;{\kern 1pt} \label{eq:subeq11d}\\
& {P_i} \le {p_{\text{BS}}},\;\;\;{\kern 1pt}  \,\,\,\,\,\,\,\, i \in \psi ,{p_{\text{BS}}} = {\rm{cte}}\label{eq:subeq11c}\\
& 0 \le {\eta _i} \le 1, \,\,\,\,\,\,\,\, i \in \psi \label{eq:subeq11e}\\
& 0 \le {\tau _i} \le 1, \,\,\,\,\,\,\,\, i \in \psi \label{eq:subeq11k}\\
& \tau _i^{{\rm{EH}}} + \tau _i^{{\rm{Bd}}} \le {\tau _i}\label{eq:subeq11f}\\
& {\varepsilon _{\text{SBD}_{i}}} \le {\Gamma} {P_i}{{\tau _i^{\text{EH}}}}|{\bf{h}}_{\text{1i}}^H{{\bf{w}}_{\text{1i}}}{|^2}, \,\,\,\,\,\,\,\, i \in \psi  \label{eq:subeq11g}\\
& R_{{\text{SUE}_{i}}}^{2,l} > R_{{\text{SUE}_{i+1}}}^{2,l},  \,\,\,\,\,\,\,\, i \in \psi, \label{eq:subeq11i}
\end{align}
\end{subequations}
where $l$ denotes the SUEs located in the transmission and reflection areas ($l=t,r$). The (\ref{eq:subeq11a}) constraint limits the power of the ASRIS to its maximum value in both transmission and reflection modes, while the constraint (\ref{eq:subeq11d}) pertains to the phase of the ASRIS elements. Additionally, constraint (\ref{eq:subeq11c}) specifies the maximum power of the BS for transmitting signals to the $\text{SBD}_{i}$, $\text{SUE}_{i}^{r}$ and ASRIS. To allocate resources more effectively, this BS power is treated as a variable in the problem, which cannot exceed the maximum specified value $p_{BS}$. Furtheremore, constraint (\ref{eq:subeq11e}) is considered to limit the rate (or power reflection) of  modulated the information on the ambient carrier by $\text{SBD}_{i}$. Constraint (\ref{eq:subeq11k}) sets the maximum duration for completing phase 1. Relation \ref{eq:subeq11f} is intended to set constraints on the energy harvesting and data transmission times for each SBD. Equation \eqref{eq:subeq11g} restricts the energy harvested by the $\text{SBD}_{i}$ to its maximum specified value. Since we assume that all SBDs are of approximately the same type in this design, we can consider the $\Gamma$ value to be the same for all of them. Also, to guarantee the SIC performed successfully, the conditions  \eqref{eq:subeq11i} should be satisfied.

It should be noted that, since SBDs merely place their own information on the carrier frequency of ambient waves, no decoding of these signals is performed. Therefore, there is no need to consider the SIC constraint for the first phase of this system model. In scenarios where decoding the BS signal is necessary, or when information exchange between SBDs in the network is required, information decoding in the SBD would also be necessary, and therefore, the SIC technique would be needed. In this case, we must consider a constraint such as
\(
{R_{\text{SBD}_{i}}^1} > {R_{\text{SBD}_{i+1}}^1},\, i \in \psi 
\)
for it. However, in this case, the SBD hardware would become more complex and might not be implementable with simple passive circuits.

To make the optimization problem (\ref{eq:main_rate}) implementable, certain modifications are required. Given the requirement of uninterrupted and collision-free transmission of information from $\text{SBD}_{i}$ to the $\text{SUE}_{i}^{l}$, the maximum achievable throughput is attained when $R_{{\rm{SB}}{{\rm{D}}_i}}^1 = R_{{\rm{SU}}{{\rm{E}}_i}}^{2,l} = R$. Therefore, the final optimization problem for maximizing the achievable rate can be stated as follows:
\begin{subequations}
\label{eq:main_rate_modification}
\begin{align}
&\max_{R,\eta_i, {{\bf{w}}_{\text{1i}}},{{\bf{w}}_{\text{2i}}}, \tau_i,\tau _i^{{\rm{EH}}},\tau _i^{{\rm{Bd}}},P_i,\beta _m^l,\theta_m^l} R \tag{15}\\
s.t. \nonumber\\
&{2^{\left( {\frac{{KR}}{{{\tau _i^{Bd}}}}} \right)}} - 1 \le \frac{{K{P_i}{\eta _i}{{\left|  {{{\bf{g}}_{\text{1i}}}} {{\bf{h}}_{\text{1i}}^H{{\bf{w}}_{\text{1i}}}} \right|}^2}}}{{\sum\limits_{j \in \upsilon } {{P_j}{\eta _j}{{\left|  {{{\bf{g}}_{\text{1j}}}}{{\bf{h}}_{\text{1j}}^H{{\bf{w}}_{\text{1j}}}} \right|}^2}}  + \sigma _{\text{BS}}^2}} \label{eq:subeq14a}\\
&{2^{\left( {\frac{R}{{1 - {\tau _i}}}} \right)}} - 1 \le \gamma _{{\rm{SU}}{{\rm{E}}_i}}^{2,r}\label{eq:subeq14c}\\
&{2^{\left( {\frac{R}{{1 - {\tau _i}}}} \right)}} - 1 \le \gamma _{{\rm{SU}}{{\rm{E}}_i}}^{2,t}\label{eq:subeq14d}\\
& (\ref{eq:subeq11a}), (\ref{eq:subeq11d}), (\ref{eq:subeq11c}), (\ref{eq:subeq11e}), (\ref{eq:subeq11k}), (\ref{eq:subeq11f}), (\ref{eq:subeq11g}), (\ref{eq:subeq11i}),
\end{align}
\end{subequations}
in problem \eqref{eq:main_rate_modification}, the constraints \eqref{eq:subeq14a}, \eqref{eq:subeq14c}, and \eqref{eq:subeq14d} correspond to the limitations imposed by the maximum information rates in a communication channel, which are defined by the relations \eqref{eq:y_R_1}, \eqref{eq:R_2 - s1r}, and \eqref{eq:R_2 - s1t}, respectively. Additionally, in the simulations for this problem, \eqref{eq:subeq14c} is used when $l=r$, and \eqref{eq:subeq14d} is used when $l=t$.

Under the aforementioned conditions, depending on factors such as the data size transmitted from the SBDs to the BS, the parameter \( K \), CSI between nodes and the BS, and ASRIS/BS beamforming coefficients, the execution times of Phase 1 and Phase 2 may be equal or different.

Problem \eqref{eq:main_rate_modification} is a non-convex problem, and it can be implemented using convex optimization methods or various machine learning techniques. Considering that resource allocation problems in various articles have mostly been simulated using convex optimization methods and the CVX toolbox in Matlab, in this paper, we will model the problem using advanced deep reinforcement learning (DRL) methods and simulate it using a python program.

It is important to note that convex optimization methods are effective for well-posed, static problems with convex objectives. However, they are less suitable for the highly complex, non-convex problem formulated in Eq.\eqref{eq:main_rate_modification}, which involves dynamic, high-dimensional decision variables such as power allocations, beamforming vectors, and reflection coefficients, along with intricate constraints. DRL, on the other hand, offers several significant advantages in this context. It is capable of handling non-convexity and large-scale action spaces, adapting to dynamic environments with uncertainty, and optimizing long-term rewards through sequential decision-making. Moreover, DRL does not require an exact closed-form model of the system, making it more robust to uncertainties and enabling effective learning in situations where convex optimization methods may fail due to the problem's inherent complexity or non-stationarity. Consequently, DRL provides a flexible and scalable solution to this optimization problem, overcoming challenges that traditional convex optimization methods would struggle to address.
\\

\section{Deep Reinforcement Learning}
In this section, we initially transform the non-convex problem \eqref{eq:main_rate_modification} into a model-free Markov decision process (MDP). Subsequently, we develop DRL algorithms utilizing PPO, TD3 and A3C to address and resolve problem \eqref{eq:main_rate_modification}.
\subsection{MDP}
The MDP formulation constructs a 4-tuple, denoted as $(\bold{s}_t, \bold{a}_t, \mathbf{r}_{t}, \bold{s}_{t+1})$, where the current state, action, reward function, and next state are represented by $\bold{s}_t$, $\bold{a}_t$, $r_{t}$, and $\bold{s}_{t+1}$, respectively. The agent interacts with the environment, observing the current state $\bold{s}_t$ from the state space $\mathcal{S}$ with $\mathbf{s}_{t}$ and selecting action $\bold{a}_t$ from the action space $\mathcal{A}$ with $\bold{a}_t$ according to its policy. Additionally, the formulation of problem \eqref{eq:main_rate_modification} further details the state, action, and reward function.
\subsubsection{State}
The current state $\mathbf{s}_t \in \mathcal{S}$ at time step $t$ encompasses crucial environmental information associated with problem \eqref{eq:main_rate_modification}. This configuration enables the policy to improve and adjust in response to the dynamic environment. To elaborate, the state $\mathbf{s}_{t}$ for the analyzed system includes the all channels of environment expressed as follows:
\begin{align} \label{state}
&\mathbf{s}_{t} =  \lbrace h_{\text{1i}}^{}, h_{2}^{}, h_{\text{3i}}^{},  g_{\text{1i}}^{},  g_{\text{2i}}^{l} \rbrace, \,\,\,\,\,\,\,\, \forall i \in \psi, \forall t \in T, l = r,t.
\end{align}
\subsubsection{Action}
The term action denoted by $\mathbf{a}_t \in \mathcal{A}$ at time step $t$ encompasses the decisions and choices undertaken by an agent as it interacts with the relevant states. These actions represent the agent's responses and strategies employed to navigate and influence the dynamics of the considered system during the specified time instance  \cite{zei2023}. Every variable within problem \eqref{eq:main_rate_modification} serves as an action, implying that each element or parameter in the problem formulation represents a distinct decision or manoeuvre in addressing and resolving the challenges posed by the problem. In essence, these variables act as the modifiable components through which the agent, system, or solver can influence and effect changes in pursuit of optimal solutions or outcomes for problem \eqref{eq:main_rate_modification}. As a result, the set of actions can be described as follows:
\begin{equation}
\begin{gathered}
  {{\mathbf{a}}_t} = \{ R,{\eta _i},{{\mathbf{w}}_{{\text{1i}}}},{{\mathbf{w}}_{{\text{2i}}}},{\tau _i},{P_i},\beta _m^l,\theta _m^l\} ,{\kern 1pt} {\kern 1pt} {\kern 1pt}  \hfill \\
    {\kern 1pt} {\kern 1pt} {\kern 1pt} {\kern 1pt} {\kern 1pt} \forall i \in \psi ,\forall m \in M,\forall t \in T,l = r,t. \hfill \\ 
\end{gathered}
\end{equation} 
\subsubsection{Reward}
The DRL methodologies instruct agents to make appropriate decisions to maximize the reward. In the optimization problem \eqref{eq:main_rate_modification}, the reward function aligns with the objective function, guiding the training process towards actions that contribute to the overall goal of optimizing the specified objective
\begin{align} 
	&\mathbf{r}_{t} =  R_{t} + \sum_{j=1}^{11} l_{C_j} , \,\,\,\,\,\,\,\, \forall t \in T,
\end{align}
where ${l_{C_j}= \alpha_{j} {R_{t}}},$ and the index $j$ corresponds to all constraints, i.e., $\forall j \in \left\{ {1,2,...,11} \right\}$. Besides, $\alpha_{j} = 1$, if the ${C_j}$-th constraint is  satisfied and $\alpha_{j} = 0$, otherwise. 

\subsection{PPO Algorithm}
The PPO algorithm, functioning as an actor-critic on-policy gradient method, simplifies the intricate computations involved in earlier policy gradient methods like trust region policy optimization (TRPO)\cite{10032267}.

In the context of reinforcement learning, the main objective is to maximize the expected cumulative reward, taking into account a protracted temporal process. Consequently, the cumulative reward at time step $t$ is represented as $\mathcal{R}_t = \sum_{t = 0}^{\infty} \lambda^{t} r_t$, where $\lambda \in \left[ {0,1} \right)$ signifies the discount factor. To delve into specifics, both actor and critic networks are employed to portray the parameterized stochastic policy of action selection, denoted as ${\pi_{\theta}}(\mathbf{a}_{t}|\mathbf{s}_{t})$, and the state-value function ${V_{\phi}(\mathbf{s}_{t})}$, respectively. Here, $\theta$ and $\phi$ stand for the parameters of the actor and critic networks. Subsequently, a surrogate objective function, constructed based on the PPO approach, can be articulated as follows:
\begin{align}
	\mathcal{L}\left( \theta, \mathbf{s}_{t}, \mathbf{a}_{t} \right) = \mathbb{E}\left[\beta_{t}(\theta) \Omega(\mathbf{s}_{t}, \mathbf{a}_{t}) \right].
\end{align}

The probability ratio between the current policy and the previous one is denoted as $\beta_{t}(\theta) = \pi_{\theta}(\mathbf{a}_{t}|\mathbf{s}_{t}) / \pi_{\theta^{\text{old}}}(\mathbf{a}_{t}|\mathbf{s}_{t})$, where $\theta^{\text{old}}$ represents the parameter associated with the old policy in the actor network. Additionally, the advantage function is expressed as follows:
\begin{align}\label{fg}
	\Omega(\mathbf{s}_{t}, \mathbf{a}_{t}) = r(\mathbf{s}_{t}, \mathbf{a}_{t}) + \lambda V_{\phi^{\text{old}}}({\mathbf{s}_{t + 1}}) - V_{\phi^{\text{old}}}({\mathbf{s}_{t}}), 
\end{align}
where $\phi^{\text{old}}$ signifies the parameter associated with the critic network for the previous state-value estimation function. Subsequently, a mini-batch stochastic gradient descent (SGD) technique is employed to update the associated $\theta$ across a set of $Q$ transitions denoted as $(\bold{s}^q_t, \bold{a}^q_t, \bold{r}^q_t, \bold{s}^q_{t+1})$ sampled from an experience buffer, which is given by
\begin{align}\label{abc}
	\theta = \theta^{\text{old}} - {\delta _A}\frac{1}{Q}\sum\limits_{q = 1}^Q {\nabla_{\theta} \tilde {\mathcal{L}}_q} \left( \theta, \mathbf{s}_{t}^{q}, \mathbf{a}_{t}^{q} \right), 
\end{align} 
where ${\delta _A}$ denotes the learning rate (LR), and $\tilde {\mathcal{L}}_q\left( \theta, \mathbf{s}_{t}^{q}, \mathbf{a}_{t}^{q} \right)$ represents the instantiation of ${\mathcal{L}}\left( \theta, \mathbf{s}_{t}, \mathbf{a}_{t} \right)$ with the $q$-th transition, respectively. The mini-batch stochastic gradient descent (SGD) utilized for updating $\phi$ employs the mean squared error (MSE) loss function in the following manner:
\begin{equation}\label{critic_update}
	\phi = \phi^{\text{old}} - \delta_C \frac{1}{Q} \sum_{q=1}^{Q} \nabla_{\phi} \big(  V_{\phi}(\mathbf{s}_{t}^{q}) - \mathcal{\hat{R}}(\mathbf{s}_{t}^{q}, \mathbf{a}_{t}^{q}) \big)^{2},
\end{equation}
where the learning rate is denoted as ${\delta _C}$. Additionally, the target state-value function, indicated as $\hat{R}(\mathbf{s}_{t}, \mathbf{a}_{t})$, is expressed as:
\begin{align}\label{target}
	\hat{R}(\mathbf{s}_{t}, \mathbf{a}_{t}) = r(\mathbf{s}_{t}, \mathbf{a}_{t}) + \lambda V_{\phi^{\text{old}}}(\mathbf{s}_{t+1}).
\end{align}

The PPO-based approach are outlined in Algorithm \eqref{algorithmPPO}. To elaborate, the action ${\mathbf{a}_t}$ is generated based on the particular policy within the current state $\mathbf{s}_{t}$, resulting in the acquisition of the reward $\bold{r}_t$. Subsequently, the transition $(\bold{s}_t, \bold{a}_t, \bold{r}_t, \bold{s}_{t+1})$ is recorded in the experience buffer, from which Q instances are sampled. In the following, the advantage function $\Omega (\mathbf{s}_{t}, \mathbf{a}_{t})$ in \eqref{fg} is computed. Finally, the relevant actor and critic parameters undergo updating through mini-batch SGD.
\begin{algorithm}
	\caption{The PPO Algorithm}\label{algorithmPPO}
		\begin{algorithmic}
		\STATE 1-	Initialize the environment parameters and the parameters \\ \,\,\,\,\,\,of actor and critic networks, i.e., $\theta$, $\phi$, $\varepsilon$, ${\delta _A}$ and ${\delta _C}$
		\STATE 2-  Set $\theta^{\text{old}}=\theta$ and $\phi^{\text{old}}=\phi$
		\STATE 3- \textbf{For} each episode do
		\STATE 4- \hspace{0.3cm} Reset environment and initialize position of users \\ \,\,\,\,\,\,\,\,\,\,\,\,\, randomly
		\STATE 5- \hspace{0.3cm} Initialize state $\mathbf{s}_{0}$ accroding to \eqref{state}
		\STATE 6- \hspace{0.3cm} \textbf{For} each step do
		\STATE 7- \hspace{0.6cm} Generate action $\mathbf{a}_{t}$ according to $\pi_{\theta}(\mathbf{a}_{t}|\mathbf{s}_{t})$ in state \\ \,\,\,\,\,\,\,\,\,\,\,\,\,\,\,\,\,\, $\mathbf{s}_{t}$
		\STATE 8- \hspace{0.6cm} Calculate reward $\mathbf{r}_{t}$ 
	    \STATE 9- \hspace{0.6cm} Observe the new state $\mathbf{s}_{t+1}$
		\STATE 10- \hspace{0.42cm} Store $(\bold{s}_t,\bold{a}_t,\bold{r}_t,\bold{s}_{t+1})$ in the experience buffer
		\STATE 11- \hspace{0.42cm} Calculate the advantage function $\Omega(\mathbf{s}_{t}, \mathbf{a}_{t})$ in \eqref{fg}
		\STATE 12- \hspace{0.42cm} Calculate ${\nabla_{\theta} \tilde {\mathcal{L}}_q} \left( \theta, \mathbf{s}_{t}^{q}, \mathbf{a}_{t}^{q} \right)$ in \eqref{abc}
		\STATE 13- \hspace{0.42cm} Calculate $\nabla_{\phi} \big(  V_{\phi}(\mathbf{s}_{t}^{q}) - \hat{R}(\mathbf{s}_{t}^{q}, \mathbf{a}_{t}^{q}) \big)^{2}$ in \eqref{critic_update}
		\STATE 14- \hspace{0.42cm} Calculate $\hat{R}(\mathbf{s}_{t}, \mathbf{a}_{t})$ in \eqref{target}
		\STATE 15- \hspace{0.42cm} Update $\theta$ and $\phi$ in  \eqref{abc} and \eqref{critic_update}, respectively
		\STATE 16-  \hspace{0.42cm} Update $\theta^{\text{old}}=\theta$ and $\phi^{\text{old}}=\phi$
		\STATE 17- \hspace{0.3cm}\textbf{End FOR}
		\STATE 18- \textbf{End FOR}
	\end{algorithmic}
\end{algorithm}

\subsection{TD3 Algorithm}
The TD3 algorithm represents a reinforcement learning approach that is both model-free and off-policy. The state-action value function as follows:
	\begin{equation}
\begin{gathered}
  {q_\mu }({{\mathbf{s}}_t},{{\mathbf{a}}_t}) =  \hfill \\
  {\mathbb{E}_{{\text{Pr}}({{\mathbf{s}}_{t + 1}}|{{\mathbf{s}}_t},{{\mathbf{a}}_t})}}[\sum\limits_{t = 0}^\infty  {{\gamma ^t}} r({{\mathbf{s}}_t},\mu ({{\mathbf{s}}_t})|{{\mathbf{s}}_t} = {{\mathbf{s}}_0},{{\mathbf{a}}_t} = \mu ({{\mathbf{s}}_0})], \hfill \\ 
\end{gathered} 
\end{equation}
where $\mu$ is parameters of the actor network, and $\gamma \in (0, 1]$ denotes the discount factor. The optimal policy is given by
	\begin{equation}
	\mu^{\star}(\mathbf{s}_{t}) = \mathrm{arg} \ \underset{\mu(\mathbf{s}_{t}) \in \mathcal{A}}{\mathrm{max}} \quad q_{\mu}(\mathbf{s}_{t}, \mu(\mathbf{s}_{t})).
\end{equation}

TD3 represents an enhanced iteration of the deep deterministic policy gradient (DDPG), introducing adjustments to mitigate the overestimation of state-action value and avert the generation of sub-optimal policies \cite{10182335}. The specifics of these modifications are elaborated in the training network description. Similar to the PPO algorithm, the agent chooses its action based on the observed state, and the corresponding experience $(\bold{s}_t, \bold{a}_t, \bold{r}_t, \bold{s}_{t+1})$ is stored in the buffer. Consequently, the loss function for critic networks with parameter $\alpha_{i}$ is calculated as follows:
\begin{equation}\label{losstd3}
	\mathcal{L}(\alpha_{i}) = \frac{1}{\lvert Q \rvert} \sum_{k=1}^{Q} \Big( q_{\mu}( \mathbf{s}_{t}^{k}, \mathbf{a}_{t}^{k}; \alpha_{i} ) - y(r_{t}^{k}, \mathbf{s}_{t+1}^{k})  \Big)^{2},
\end{equation}
where y is calculated by
\begin{equation} 
	y(r_{l}^{k}, \mathbf{s}_{t+1}^{k}) = r_{t}^{k} + \gamma \underset{i}{\mathrm{min}} q_{\bar{\mu}} (\mathbf{s}_{t+1}^{k}, \tilde{\mathbf{a}}_{t+1}^{k}; \bar{\alpha}_{i}).
\end{equation}

To adjust the parameters of critic networks, $\alpha_{i}$, the gradient descent algorithm is employed on the loss function \eqref{losstd3} using the following equation:
\begin{equation}\label{parameters_of_TD3_critics}
	\alpha_{i} = \alpha_{i} - \theta_{i} \nabla_{\alpha_{i}} \mathcal{L}(\alpha_{i}),
\end{equation}
where $\theta_{i}$ represents the learning rate. Also, the loss function and update parameters of the actor network are as follows:
\begin{equation}
	\mathcal{L}(\mu) = \frac{1}{\lvert Q \rvert} \sum_{k=1}^{Q} q_{\mu}( \mathbf{s}_{l}^{k}, \mathbf{a}_{l}^{k}),
\end{equation}
\begin{equation}\label{parameter_TD3_AN}
	\mu = \mu - \tilde \theta \nabla_{\mu} \mathcal{L}(\mu).
\end{equation}

The pseudo-code for the TD3 algorithm is presented in Algorithm \eqref{algorithmTD3}.

\begin{algorithm}
	\caption{The TD3 Algorithm}\label{algorithmTD3}
	\begin{algorithmic}
		\STATE 1-	Initialize the environment and networks parameters, i.e., \\ \,\,\,\,\, $\alpha$, and $\mu$
		\STATE 2-  Set $\theta^{\text{old}}=\theta$ and $\phi^{\text{old}}=\phi$
		\STATE 3- \textbf{For} each episode do
		\STATE 4- \hspace{0.3cm} Reset environment and initialize position of users \\ \,\,\,\,\,\,\,\,\,\,\,\,\,  randomly
		\STATE 5- \hspace{0.3cm} Initialize state $\mathbf{s}_{0}$ accroding to \eqref{state}
		\STATE 6- \hspace{0.3cm} \textbf{For} each step do
		\STATE 7- \hspace{0.6cm} Observe state $\mathbf{s}_{t}$ and select action $\mathbf{a}_{t}$
		\STATE 8- \hspace{0.6cm} Observe next state $\mathbf{s}_{l+1}$ and receive reward $r_{t}$ 
		\STATE 9- \hspace{0.6cm} Store $\left(\mathbf{s}_t, \mathbf{a}_t, r_t, \mathbf{s}_{t+1} \right)$ in $\mathcal{M}$.
		\STATE 10- \hspace{0.42cm} Randomly sample a batch set $B$ from reply buffer
		\STATE 11- \hspace{0.42cm} Update the network parameters $\alpha_{i}$, and $\mu$ using \eqref{parameters_of_TD3_critics}, \\ \hspace{1cm}\eqref{parameter_TD3_AN}, respectively. 
		\STATE 12- \hspace{0.3cm}\textbf{End FOR}
		\STATE 13- \textbf{End FOR}
	\end{algorithmic}
\end{algorithm}

\subsection{A3C Algorithm}
The A3C algorithm employs an actor-critic architecture, where an actor network makes policy decisions and a critic network evaluates the value of these decisions. The "Advantage" in A3C refers to the use of an advantage function, which measures the advantage of taking a particular action in a given state over the average action value \cite{9385791}. the advantage function is used to decrease the variance in estimation, as expressed by the following equation:
\begin{equation}
	\mathcal{A}_t(\mathbf{s}_t, \mathbf{r}_t; \mu, \alpha) = \mathcal{R}_t - V(s_t;\alpha),
\end{equation}
where $\mu$ and $\alpha$ are the parameters of actor and critic network, respectively. Furthermore, $\mathcal{R}$ represents cumulative reward as follows:
\begin{equation}
\mathcal{R} = \sum_{i=0}^k \gamma^i r_{t+i} + \gamma^k V (s_{t+k};\alpha),
\end{equation}
where k is the number of steps which A3C used for parameter updating.

Derived from the advantage function $\mathcal{A}_t$ the actor's loss function is expressed as
\begin{equation}
	f_{\pi}(\mu) = \log \pi(a_t | s_t; \mu)( \mathcal{A}_t) + \beta H(\pi (s_t;\mu)),
\end{equation}

Here, $H(\pi (s_t;\mu))$ represents an entropy term incorporated to promote exploration during training, preventing potential premature convergence \cite{CHEN2020485}. The parameter $\beta$ is utilized to regulate the intensity of entropy regularization, facilitating the balance between exploration and exploitation. The loss function for the estimated critic network is specified as:
\begin{equation}
	f(\alpha) = (\mathcal{R}_t - V(s_t;\alpha))^2.
\end{equation}

This is employed to update the value function \(V(st; \theta_v)\). The update for the critic is executed using the following accumulated gradient:
\begin{equation} \label{dalpha}
	d\alpha \leftarrow 	d\alpha + \frac{\partial(\mathcal{R}_t - V(s_t;\alpha))^2}{\partial \alpha^{\prime}}.
\end{equation}

The actor undergoes an update through the following process:
\begin{equation} \label{dmu}
	d\mu \leftarrow d\mu + \nabla_{\mu^\prime} \log \pi(a_t | s_t; \mu^\prime)( \mathcal{A}_t) + \beta \nabla_{\mu^\prime} H(\pi (s_t;\mu^\prime)).
\end{equation}

All the steps of the A3C algorithm are outlined in Algorithm \eqref{algorithmA3C}.

\begin{algorithm}
	\caption{The A3C Algorithm}\label{algorithmA3C}
	\begin{algorithmic}
		\STATE 1-  Initialize the global actor network and global critic \\ \,\,\,\,\,\, network with
		parameters, $\alpha$, and $\mu$
		\STATE 2-  Initialize the thread-specific actor and thread-specific \\ \,\,\,\,\,\, critic network parameters $\alpha^\prime$, and $\mu^\prime$
		\STATE 3- \textbf{For} each episode do
		\STATE 4- \hspace{0.3cm} Reset environment and initialize position of users\\ \,\,\,\,\,\,\,\,\,\,\,\,\, randomly
		\STATE 5- \hspace{0.3cm} \textbf{For} each worker do
		\STATE 6- \hspace{0.6cm} Initialize the gradients of global agent: $d\alpha = 0$, \\ \,\,\,\,\,\,\,\,\,\,\,\,\,\,\,\,\,\, $d\mu = 0$
		\STATE 7- \hspace{0.6cm} Synchronous parameters of each worker with global  \\ \,\,\,\,\,\,\,\,\,\,\,\,\,\,\,\,\,  parameters  $\alpha^\prime = \alpha$, and $\mu^\prime = \mu$
		\STATE 6- \hspace{0.6cm} Obtain initial state $s_0$.
		\STATE 7- \hspace{0.75cm} \textbf{For} each step do
		\STATE 8- \hspace{1cm} Perform at under policy $\pi (s_t;\mu^\prime)$
		\STATE 9- \hspace{01cm} Obtain reward $r_t$ and new state $s_{t+1}$
		\STATE 10- \hspace{0.6cm} \textbf{End FOR}
		\STATE 11- \hspace{0.3cm} $R = \begin{cases} 0, \, \, \, \, \, \, $for terminal state$ \\ V(s_t;\alpha^\prime), \, \, \, \, \, \, $for non-terminal state$  \end{cases}$
		\STATE 12- \hspace{0.3cm} $R = r_t +  \gamma R$
		\STATE 13- \hspace{0.3cm} Obtain accumulate gradient $\alpha$ based on \eqref{dalpha}
		\STATE 14- \hspace{0.3cm} Obtain accumulate gradient $\mu$ based on \eqref{dmu}
		\STATE 15- \hspace{0.25cm} \textbf{End FOR}
		\STATE 16- \textbf{End FOR}
	\end{algorithmic}
\end{algorithm}

In the following, each of the PPO, TD3, and A3C methods, modeling and simulations are conducted, and the outputs of each are compared with each other in relation to this modeling system.

\section{Simulation Results}

According to the proposed model, SBDs and SUEs are randomly distributed across the network space, as depicted in Fig. \ref{fig:STAR_RIS_scheme 4}. In all simulations, constant values of \(K = 100\), Rician factor = 10, $\Gamma=0.8$, and \(\sigma _{\text{ASRIS}}^2 = \sigma _{\text{BS}}^2 = \sigma _{\text{SUE}}^2 = -120 \, \text{dBm}\) have been considered. Additionally, we assume the carrier frequency of the ambient signal is 28 GHz, the path loss exponent is 3, and the transmit power of the BS and ASRIS are 16 watts and 8 watts, respectively. All simulation results are generated by averaging over 100 random channels. The maximum distance between the BS and SBDs is approximately 200 meters, while the distance between the BS and ASRIS extends up to about 400 meters. Within the ASRIS transmission area, the distance from the ASRIS to the SUEs is 300 meters. SUEs in the Reflection area are randomly distributed within the designated space between the BS and ASRIS.

The simulations were carried out on a laptop featuring an Intel Core i7-6500U 8 GB DDR3-RAM.
 It should be noted that the parameters related to learning simulation for each method are given in Table   \ref{table:learning parameter}.

\begin{table}
    \centering
    \caption{Parameters set in the learning simulation.}
    \begin{tabular}{|c|c|c|c|}
        \Xhline{1.2pt} 
        \textbf{Parameters} & \textbf{PPO} & \textbf{TD3} & \textbf{A3C}\\
        \Xhline{1.2pt} 
        Number/Size of Actor,Critic & 2/128,128 & 2/400,300 & 2/128,128 \\
        \hline
        Min Batch Size & 32 & 64 & 64 \\
        \hline
        Actor Learning Rate & 0.0001 & 0.0001  & 0.0001  \\
        \hline
        Critic Learning Rate & 0.001  & 0.001  & 0.001  \\
        \hline
        Target Network Update & 0.0005 & 0.0005 & 0.0005 \\
        \hline
        Discount Factor & 0.99 & 0.99 & 0.99 \\
        \hline
        Policy Entropy Coefficient & 0.01 & - & - \\
        \hline
        Number of Workers & - & - & 3 \\
        \hline
        Number of Episodes & 30000 & 30000 & 30000 \\
        \hline
        Number of Steps & 200 & 200 & 200 \\
        \Xhline{1.2pt}
    \end{tabular}
    \label{table:learning parameter}
\end{table}

 \subsection{Convergence of the Proposed Methods}
 As described, in this paper, we utilized three learning methods, namely PPO, TD3, and A3C, to tackle the non convex optimization problem. In this scenario, the convergence plot for each of these methods is depicted in Fig. \ref{fig:reward}.

\begin{figure}
\centerline{\includegraphics[width=8 cm]{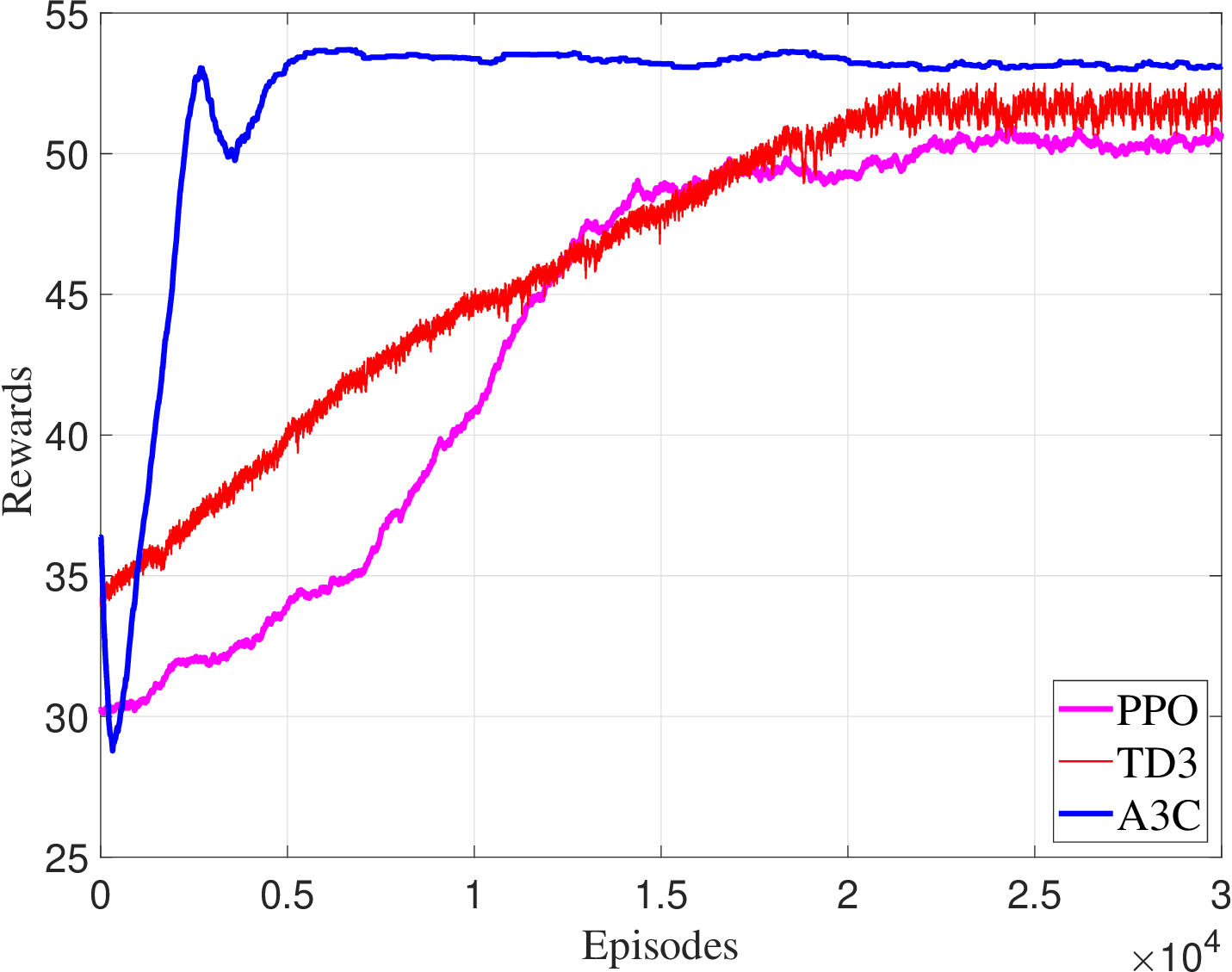}}
\caption{Convergence plot of the three learning methods, PPO, TD3, and A3C, in the case of \(
N = 8, M = 16, I = 3, \varepsilon_{\text{SBD}_i} = 1 \,\, \si{\micro\watt}
\).}
\label{fig:reward}
\end{figure}

Based on the information presented in Fig. \ref{fig:reward}, it is clear that the A3C method exhibits markedly superior convergence compared to the other two methods. Specifically, the convergence points for PPO, TD3, and A3C occur at approximately 17000, 22000, and 5000 episodes, respectively. These values indicate when network learning completes for each method. Furthermore, their convergence demonstrates the validity of their performance on this network.

\subsection{Throughput Maximization Based on Energy Harvesting by SBDs}

In the design of passive symbiotic radio networks, a pivotal factor is the quantity of energy harvested by the SBD devices. In this section, adhering to the constraint outlined in \eqref{eq:subeq11g}, we depict the rate of information in the network as it correlates with the fluctuations in the energy harvesting capacity for each SBD.

\begin{figure}
\centerline{\includegraphics[width=8 cm]{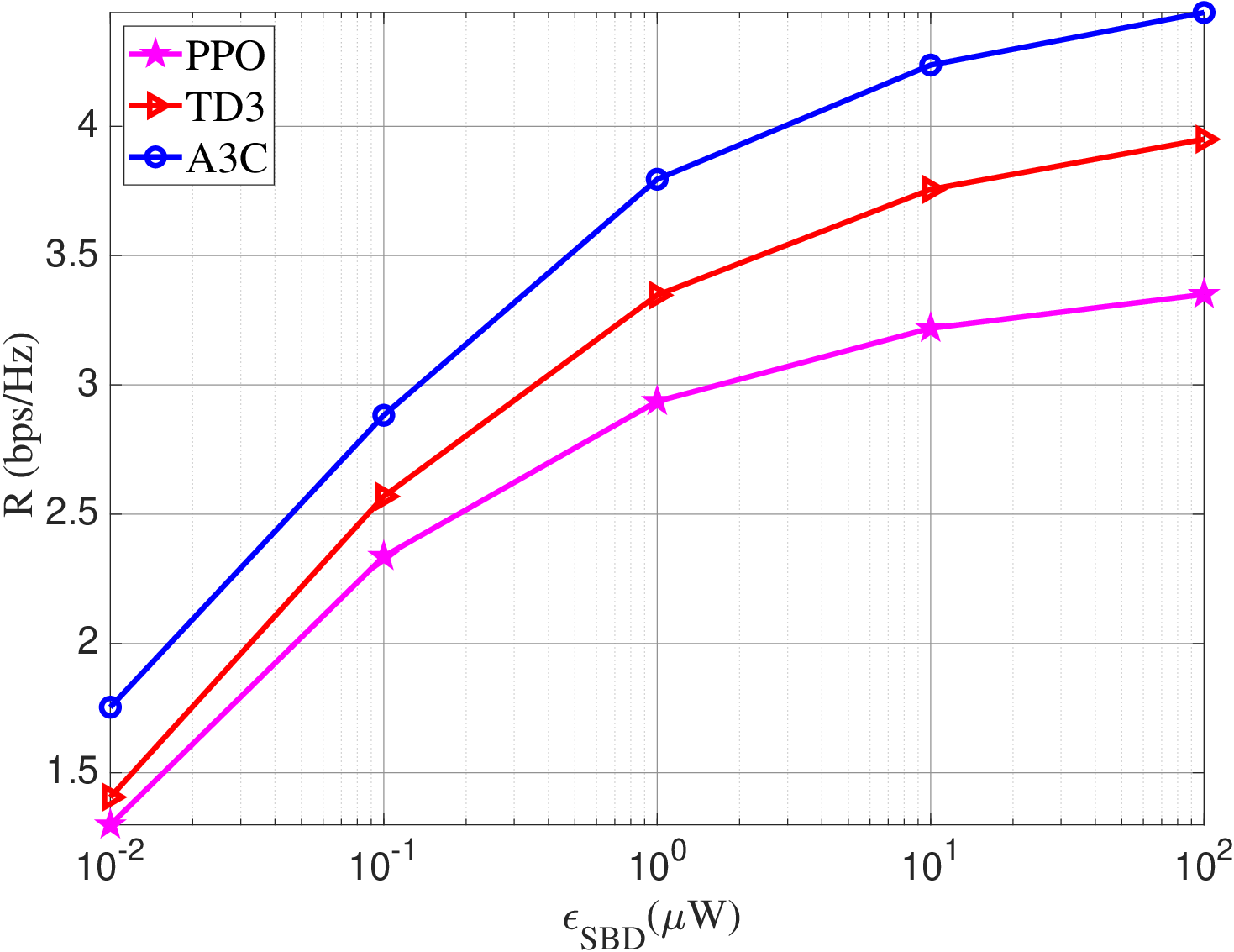}}
\caption{Throughput maximization based on the energy harvested by SBDs in all three methods, PPO, TD3, and A3C, under the conditions \(N = 8, M = 16, I = 3\).}
\label{fig:R_epsilon}
\end{figure}

As illustrated in Fig. \ref{fig:R_epsilon}, the information throughput in the network increases with the energy harvesting capability of the devices. This rise in throughput is due to the enhanced ability of each SBD to store energy from ambient waves, which in turn accelerates the speed of information modulation. Consequently, the first phase rate, linked to the CSR structure, also experiences an increase.
It is worth noting that the average energy required for transmitting a pulse by passive IoT devices is about 1-10 \si{\micro\watt} \cite{r22222,r6333,r6444,r6555}.

Moreover, in this approach, it is apparent that utilizing the A3C algorithm leads to an increase in the total information exchange rate compared to the other two methods.

\subsection{Throughput Maximization versus Transmit Power of BS and ASRIS}

Since both BS and STAR-RIS are considered active in the proposed system model, in this section, we investigate the impact of the transmit power of each on the network's performance.

In Fig. \ref{fig:R_powerBS}, we investigate the influence of augmenting the transmit power of the BS from 4 to 32 watts across all three methods. Additionally, Fig. \ref{fig:R_powerIRS} illustrates the consequences of amplifying the transmit power of the ASRIS from 2 to 16 watts,
which is caused by the reflection and transmission of the signals by it. The analysis encompasses all three methods, namely PPO, TD3, and A3C.

\begin{figure}
\centerline{\includegraphics[width=8 cm]{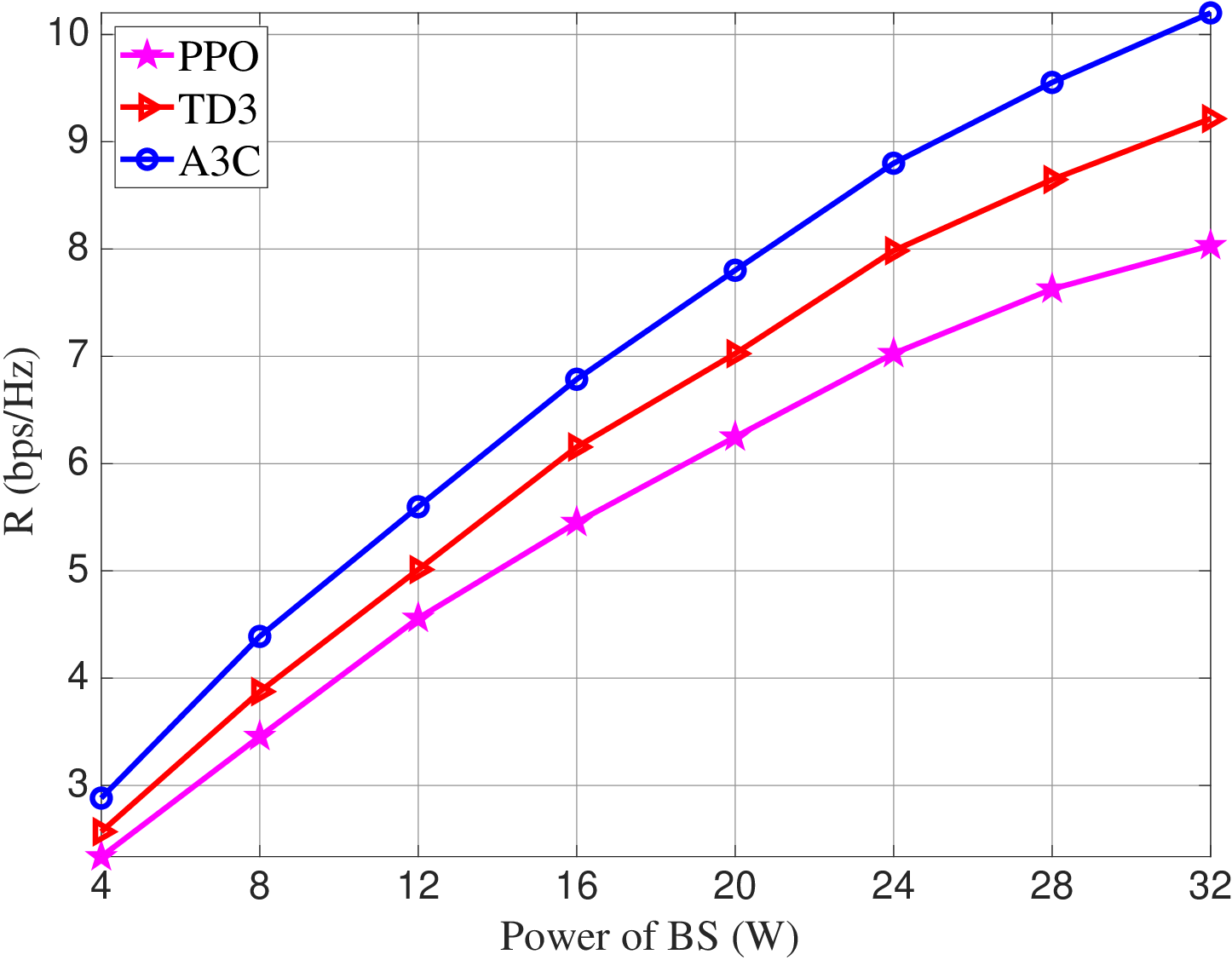}}
\caption{Impact of increasing the transmitted signal power at the BS in relation to the user data rate in the case of \(N = 8, M = 16, I = 3, \varepsilon _{\text{SBD}_i} = 1 \,\, \si{\micro\watt}\), $P_{\text{ASRIS}}=10 \,\, W$.}
\label{fig:R_powerBS}
\end{figure}\begin{figure}
\centerline{\includegraphics[width=8 cm]{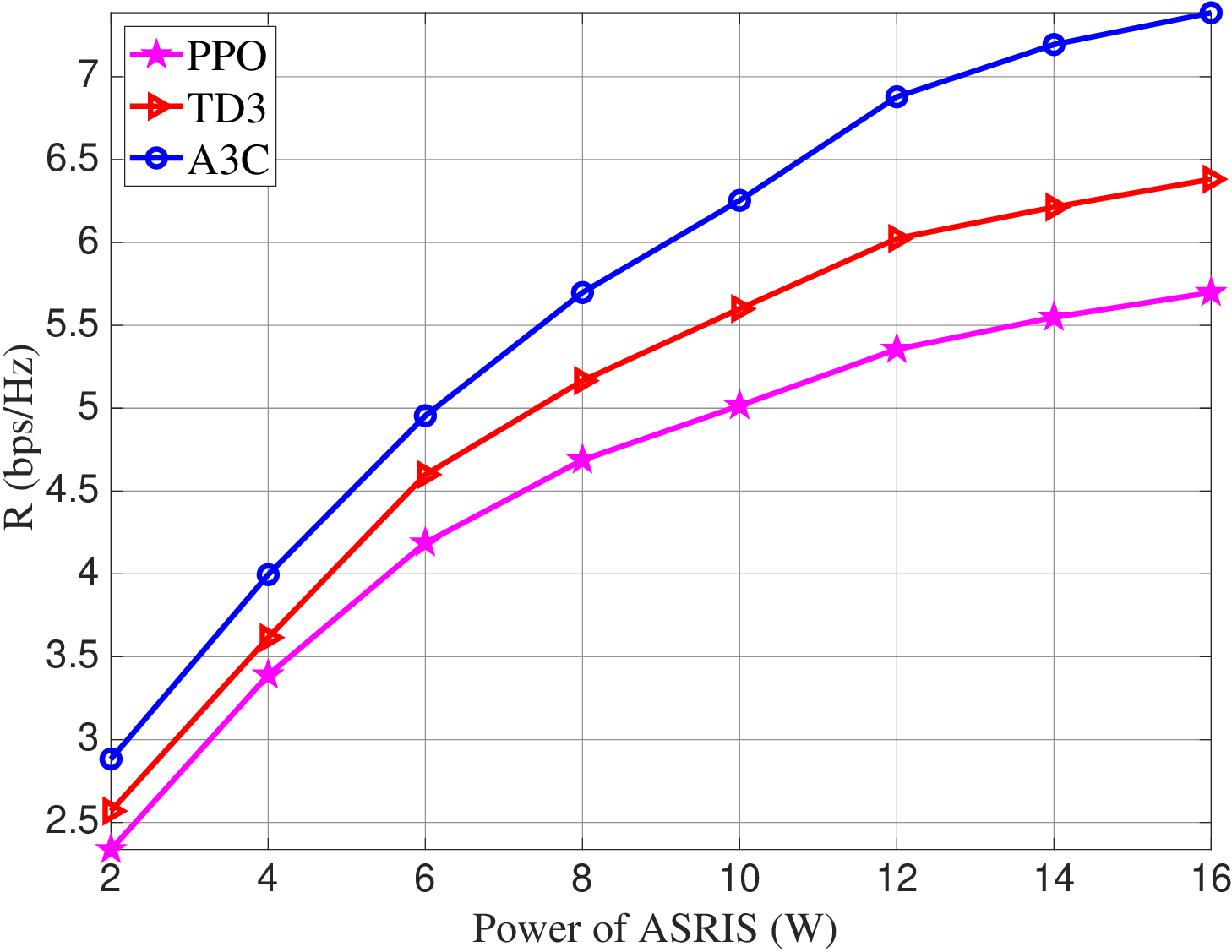}}
\caption{Impact of increasing the transmitted signal power in ASRIS in relation to the user data rate in the case of \(N = 8, M = 16, I = 3, {\varepsilon _{\text{SBD}_i}} = 1 \,\, \si{\micro\watt}\), $P_{\text{BS}}=20 \,\, W$.}
\label{fig:R_powerIRS}
\end{figure}

As expected, with the increase in the transmit power of signals in each BS or ASRIS, the data rate also increases. According to Figs. \ref{fig:R_powerBS} and \ref{fig:R_powerIRS} , the A3C, TD3, and PPO methods allocate the highest processing efficiency for optimal resource allocation among users, consequently increasing the users' information rate. Also, considering the convergence depicted in Fig. \ref{fig:reward}, indicating better convergence of the A3C method compared to the other two methods, we conclude that this method is more suitable for modeling and implementing the proposed system in this research.

\subsection{Throughput Maximization versus the Number of Elements in BS and ASRIS}

As mentioned in the first part of this chapter, our system model incorporates a BS equipped with Massive MIMO antennas and an active STAR-RIS within the network. In this section, our objective is to examine the influence on the data rate of all users in the network by manipulating the number of elements in both the BS and ASRIS.

\begin{figure*}[]
    \centering
    \begin{minipage}[b]{.32\textwidth}
        \centering
        \includegraphics[width=\textwidth]{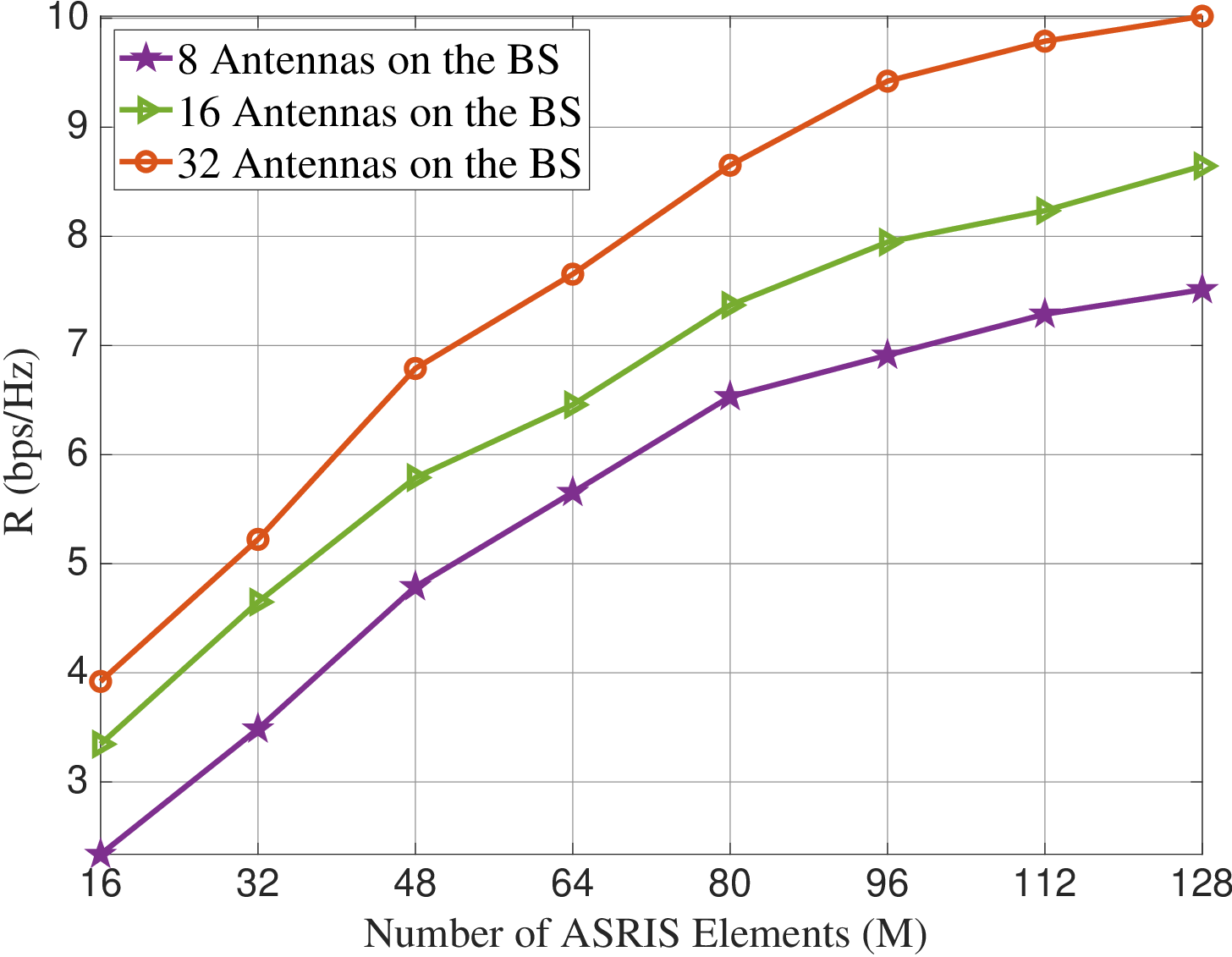}
         \caption*{\textbf{(a)}}
        \label{subfig:ppo}
    \end{minipage}
    \hfill
    \begin{minipage}[b]{.32\textwidth}
        \centering
        \includegraphics[width=\textwidth]{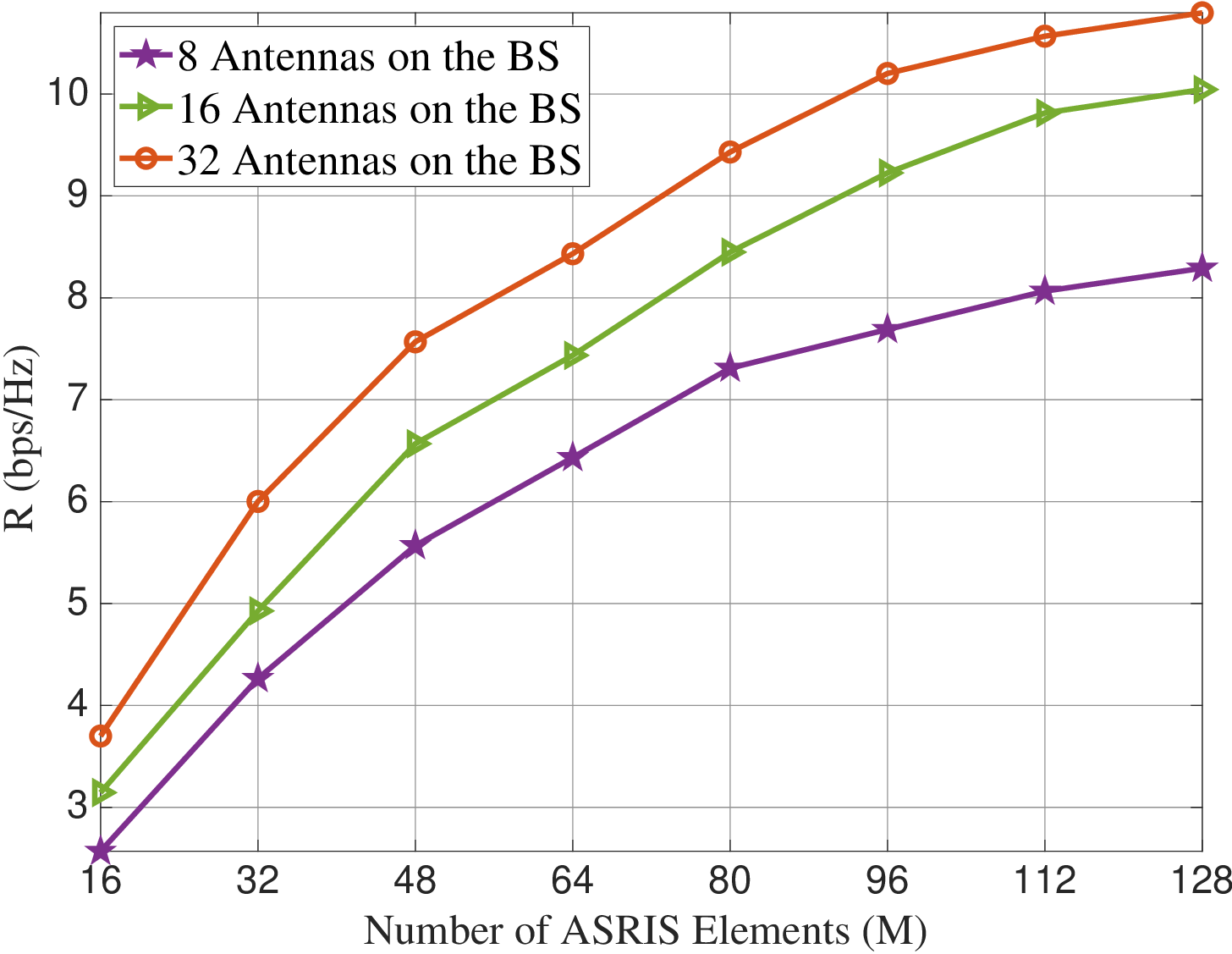}
         \caption*{\textbf{(b)}}
        \label{subfig:td3}
    \end{minipage}
    \hfill
    \begin{minipage}[b]{.32\textwidth}
        \centering
        \includegraphics[width=\textwidth]{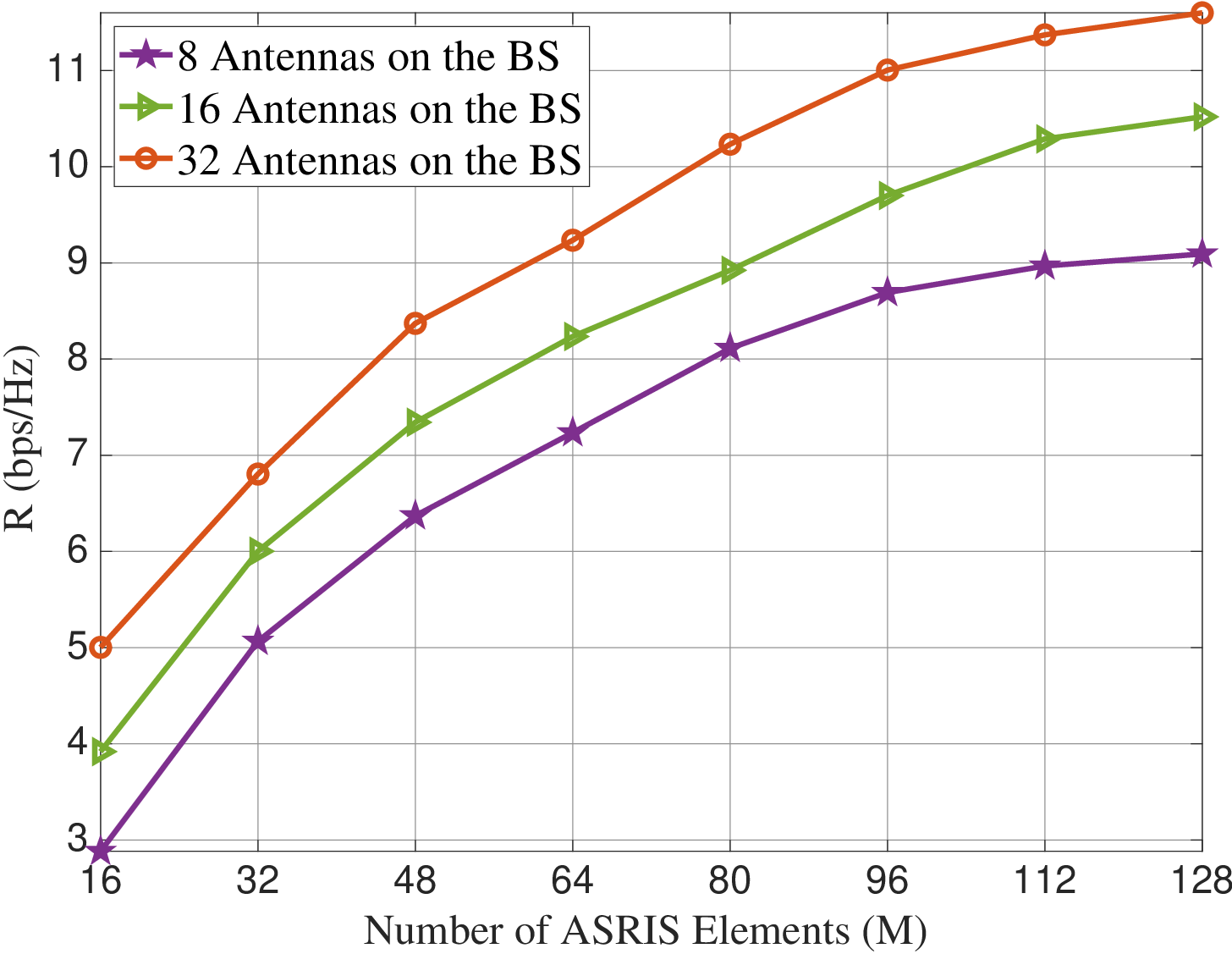}
         \caption*{\textbf{(c)}}
        \label{subfig:a3c}
    \end{minipage}
    \caption{Throughput maximization versus varying the number of elements in BS and ASRIS in \textbf{(a)} PPO, \textbf{(b)} TD3, and  \textbf{(c)} A3C methods, with \(I = 3, {\varepsilon _{\text{SBD}_i}} = 1\,\, \si{\micro\watt}\).}
    \label{fig:ppotd3a3c}
\end{figure*}

As illustrated in Fig. \ref{fig:ppotd3a3c}-(a), which belongs to the PPO method, increasing the number of elements in ASRIS from 16 to 128 exhibits a notable upward slope in the total data exchange rate. This phenomenon is attributed to more precise beamforming allocation and higher power for each of the SUEs. On the other hand, by augmenting the number of elements in BS from 8 to 32, the network's performance sees a significant improvement of approximately 133\%. Consequently, users can engage in information exchange at a much higher rate compared to the previous rate In this scenario, the cumulative data exchange rate can achieve 10 bps/Hz.

An noteworthy consideration in this scenario is the increase in the volume of information processing in the network, which amplifies the computational complexity and implementation challenges as the number of elements grows. Therefore, careful consideration must be given to determining an optimal number of elements.

Now, in light of the aforementioned details, for better comparison between these methods, we plot the graphs for the TD3 and A3C methods.

As evident in Figs. \ref{fig:ppotd3a3c}-(b), and \ref{fig:ppotd3a3c}-(c), the data rate in the A3C method is, on average under similar conditions, higher at 1 bps/Hz compared to the TD3 method. This scale holds approximately true for the TD3 graphs in comparison to the PPO ones as well.

\subsection{Comparison of Throughput Maximization versus the Number of Elements in Active  and Passive STAR-RIS}

In this section, we aim to assess the efficacy of activating a STAR-RIS as opposed to its passive counterpart. In the optimization problem \eqref{eq:main_rate_modification} considered in the system model, we have constraint \eqref{eq:subeq11c}, corresponding to the ASRIS mode in the network. To examine and compare the ASRIS structure with the passive STAR-RIS mode, we need to replace constraint \eqref{eq:subeq11c} with the relationship $\beta_{m}^{t} + \beta_{m}^{r} = 1$.

\begin{figure}
\centerline{\includegraphics[width=8 cm]{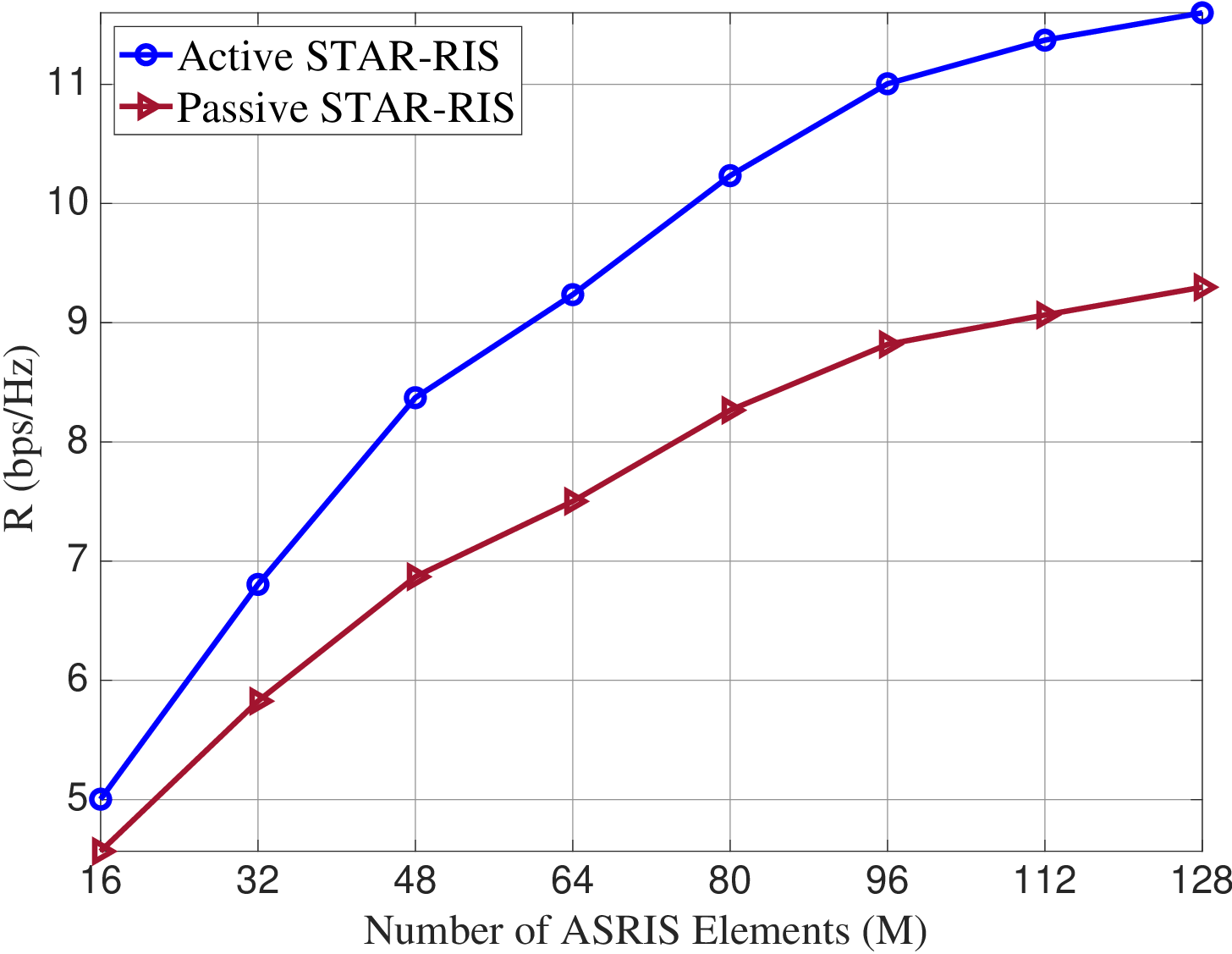}}
\caption{Throughput maximization versus the number of elements in active and  passive STAR-RIS in A3C method.}
\label{fig:R_elements_A3C_passive}
\end{figure}

As depicted in Fig. \ref{fig:R_elements_A3C_passive}, the information throughput is almost 2 bps/Hz higher when actively employing STAR-RIS compared to the passive utilization of STAR-RIS. This value further increases with the addition of more elements to the STAR-RIS configuration. Furthermore, it is evident that an ASRIS structure with 64 elements achieves the same efficiency as a passive structure with 128 elements. Put differently, the desired performance can be achieved with half the number of elements in this configuration. Therefore, reducing the number of elements in the active RIS within the network can significantly decrease implementation complexity and facilitate the estimation of communication channels in complex urban environments.

\subsection{Comparison of Proposed Model with Other Baselines}

To comprehensively evaluate the proposed scheme, we compare four different structures in this section:
\begin{enumerate}
\item{ ASRIS with 64 and 128 elements}
\item{ Passive STAR-RIS with 64 and 128 elements}
\item{ Random phase selection for RIS (without optimization)}
\item{ Replacing RIS with several passive relays}
\end{enumerate}

For a fair comparison of the above schemes, we consider users on only one side since a relay cannot simultaneously serve users on both its front and back. Therefore, in this section, the STAR-RIS is converted to the RIS structure.

The baseline 3) indicates that the RIS parameters have not been optimized by the agent. Instead, we use randomly generated $\theta_m$ values during the simulation. By adopting this approach, we can assess the performance of the system without the influence of optimized parameters and compare it with more sophisticated strategies.
In the fourth baseline, we assume that instead of using RIS in the network, several passive relays are employed to transmit information and enhance diversity to the destination. This method is discussed in paper \cite{gong2018backscatter}. Each of these relays has a structure similar to the SBDs used in the first phase, thereby creating a backscatter relay configuration within a cellular network.

\begin{figure}
\centerline{\includegraphics[width=8 cm]{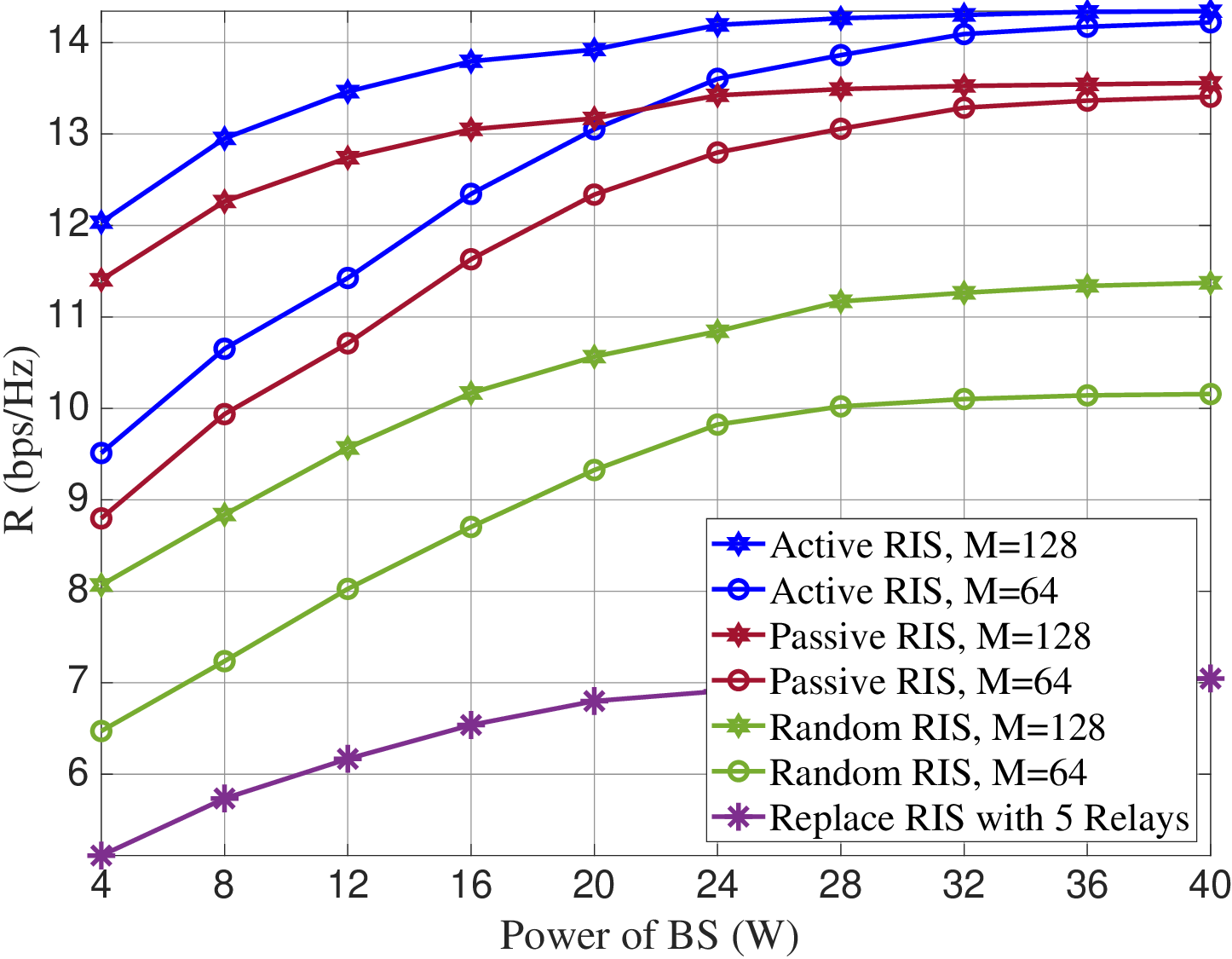}}
\caption{Comparison of active, passive, random phase RIS states, and their replacement with 5 passive relays, \(N = 8, I = 3, P_{\text{ASRIS}}=10 \,\, W\).}
\label{fig:R_relay_comparision}
\end{figure}

We mathematically model all four of the aforementioned cases and simulate them using the A3C method, resulting in Fig. \ref{fig:R_relay_comparision}. As depicted in the figure, the proposed method in this paper demonstrates a significant advantage over other methods. In this configuration, using five relays not only fails to achieve optimal efficiency but also introduces considerable complexity to the network. It is noteworthy that increasing the number of relays makes it impossible to eliminate the interference they cause at the destination.

Another significant observation is the convergence of the 128 and 64 element diagrams in both active and passive modes of the RIS. This suggests that as the transmission power from the BS increases, it reaches a saturation point beyond which further increments do not substantially enhance the total network rate. Additionally, augmenting the number of RIS elements also ceases to significantly boost the overall rate beyond a specific threshold.
On the other hand, comparing the Random phase diagrams with other diagrams where RIS phase optimization has been performed clearly shows that phase optimization in the network is essential for implementing telecommunication networks.

\subsection{Power Consumption Comparison: passive STAR-RIS vs. ASRIS}

We compare the power consumption of passive STAR-RIS and ASRIS with a 16-element structure. For passive STAR-RIS, each element employs a passive phase-shifting control circuit with a power consumption of approximately $6 \, \text{mW}$ (The power consumption  of each phase shifter, which are 1.5, 4.5, 6, and 7.8 mW for resolutions of 3, 4, 5, and 6 bits, respectively \cite{huang2019}). Additionally, a switching circuit required for dual-mode operation (reflection and transmission) consumes about $1 \, \text{mW}$ per element \cite{long2021active}. Therefore, the total energy consumption for passive STAR-RIS is calculated as $P_{\text{total, passive STAR-RIS}} = M \times \left( P_{\text{control}} + P_{\text{switching}} \right)$, where $M = 16$. Substituting the values yields $P_{\text{total, passive STAR-RIS}} = 16 \times \left( 6 \, \text{mW} + 1 \, \text{mW} \right) = 112 \, \text{mW}$.

For ASRIS, each element has a control circuit and switching circuit with the same power consumption as STAR-RIS, but also includes an active amplifier to boost the transmitted or reflected signal. The power consumption of the amplifier ranges from $50 \, \text{mW}$ to $100 \, \text{mW}$, with an average of $75 \, \text{mW}$ considered in this analysis \cite{long2021active}. Thus, the total energy consumption for ASRIS is given by $P_{\text{total, ASRIS}} = M \times \left( P_{\text{control}} + P_{\text{switching}} + P_{\text{amplifier}} \right)$. Substituting the values gives $P_{\text{total, ASRIS}} = 16 \times \left( 6 \, \text{mW} + 1 \, \text{mW} + 75 \, \text{mW} \right) = 1312 \, \text{mW}$.

As can be observed, the energy consumption of ASRIS is significantly higher than that of STAR-RIS. Specifically, ASRIS consumes approximately $11.7$ times more energy than passive STAR-RIS for a 16-element structure. Factors influencing ASRIS energy consumption include amplifier gain, number of elements, and material efficiency.
\\
\section{Conclusion}

In this article, we have explored a comprehensive system model incorporating various cutting-edge technologies, including massive MIMO, ASRIS, and both passive and active users. The objective of this initiative is to achieve optimal resource allocation among users, aiming to maximize the throughput across the entire network. In this system, for a practical modeling approach, we have incorporated constraints to ensure the minimum required QoS, impose limits on the maximum power for both BS and ASRIS, account for the constraints on the amount of energy harvesting for SBDs, and adhere to the requirements of satisfying SIC in the NOMA multiple access scheme. After mathematically modeling the target system, we encounter a non-convex and complex problem. To address the objectives inherent in this challenge, we leverage advanced DRL methods, including PPO, TD3, and A3C. The conducted simulations reveal that the A3C method, apart from achieving faster convergence, exhibits the capability to enhance the total throughput rate in the network when compared to the other two methods, TD3 and PPO. Additionally, the TD3 method significantly outperforms the PPO method. In the final segment of the simulation section, we draw the conclusion that the adoption of an active structure significantly influences the network throughput rate in comparison to the passive STAR-RIS mode. Additionally, in this section, we observed that using RIS in the network is significantly more efficient than using the backscatter relay structure. Moreover, with the help of the provided diagrams, we can determine the optimal network performance point for implementation by selecting the optimal number of RIS elements and the transmission power from the BS.

\bibliographystyle{IEEEtran}
\bibliography{StarRIS}

\begin{IEEEbiography}[{\includegraphics[width=1in,height=1.25in,clip,keepaspectratio]{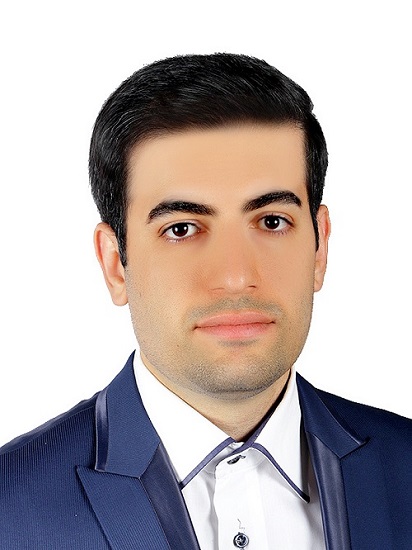}}]{Rahman Saadat Yeganeh}
received his Ph.D. degree in Electrical Engineering (Telecommunication Systems) from Isfahan University of Technology (IUT) in 2024. He is currently a Postdoctoral Researcher in Telecommunication Systems at Sharif University of Technology (SUT), Tehran, Iran. Beyond academia, Dr. Saadat is actively involved in the research and design of various telecommunication systems. His research interests include wireless communication systems with a particular focus on 5G and 6G cellular networks and IoT, as well as satellite communications, symbiotic radio, reconfigurable intelligent surfaces (RIS), non-orthogonal multiple access (NOMA), and signal processing.
\end{IEEEbiography}
%\vspace{-1.2cm}
\begin{IEEEbiography}[{\includegraphics[width=1in,height=1.25in,clip,keepaspectratio]{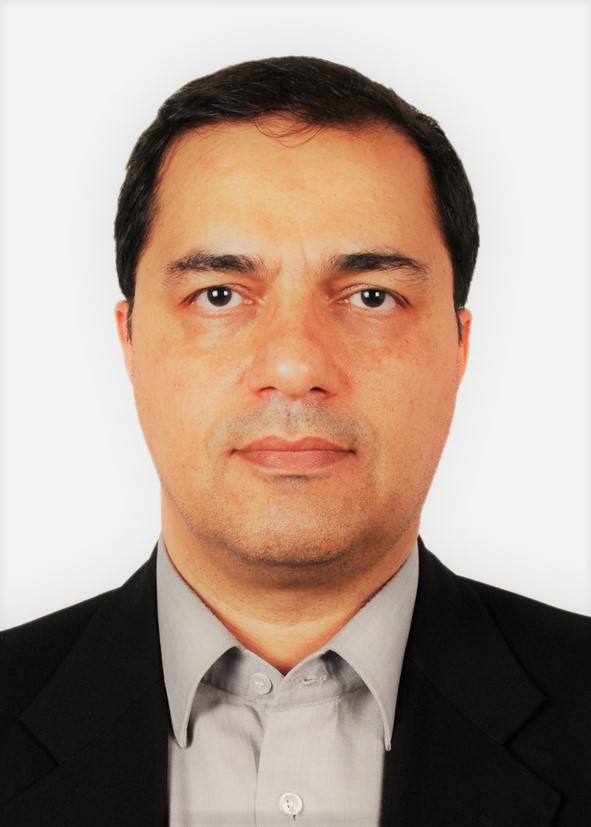}}]{Mohammad Javad Omidi}
received his Ph.D. from the University of Toronto in 1998. He has extensive industry experience in Canada, specializing in the design of broadband communication systems.

He has held several prominent academic and administrative positions at Isfahan University of Technology (IUT) in Iran, including Professor in the Department of Electrical and Computer Engineering (ECE), Chair of the IT Center, Chair of the ECE Department, and Vice President for Research and Technology. Currently, he serves as a Professor in the ECE Department at Kuwait College of Science and Technology (KCST).

His research focuses on Wireless Communications, Digital Communication Systems, and Cognitive Radio Systems. He has authored numerous publications and holds six U.S. patents along with four international patents in these areas.

Beyond academia, Prof. Omidi has played a significant role in fostering innovation and entrepreneurship. He has extensive experience in managing science parks, supporting technology start-ups, and mentoring university graduates in their entrepreneurial ventures.
\end{IEEEbiography}
%\vspace{-1.2cm}
\begin{IEEEbiography}[{\includegraphics[width=1in,height=1.25in,clip,keepaspectratio]{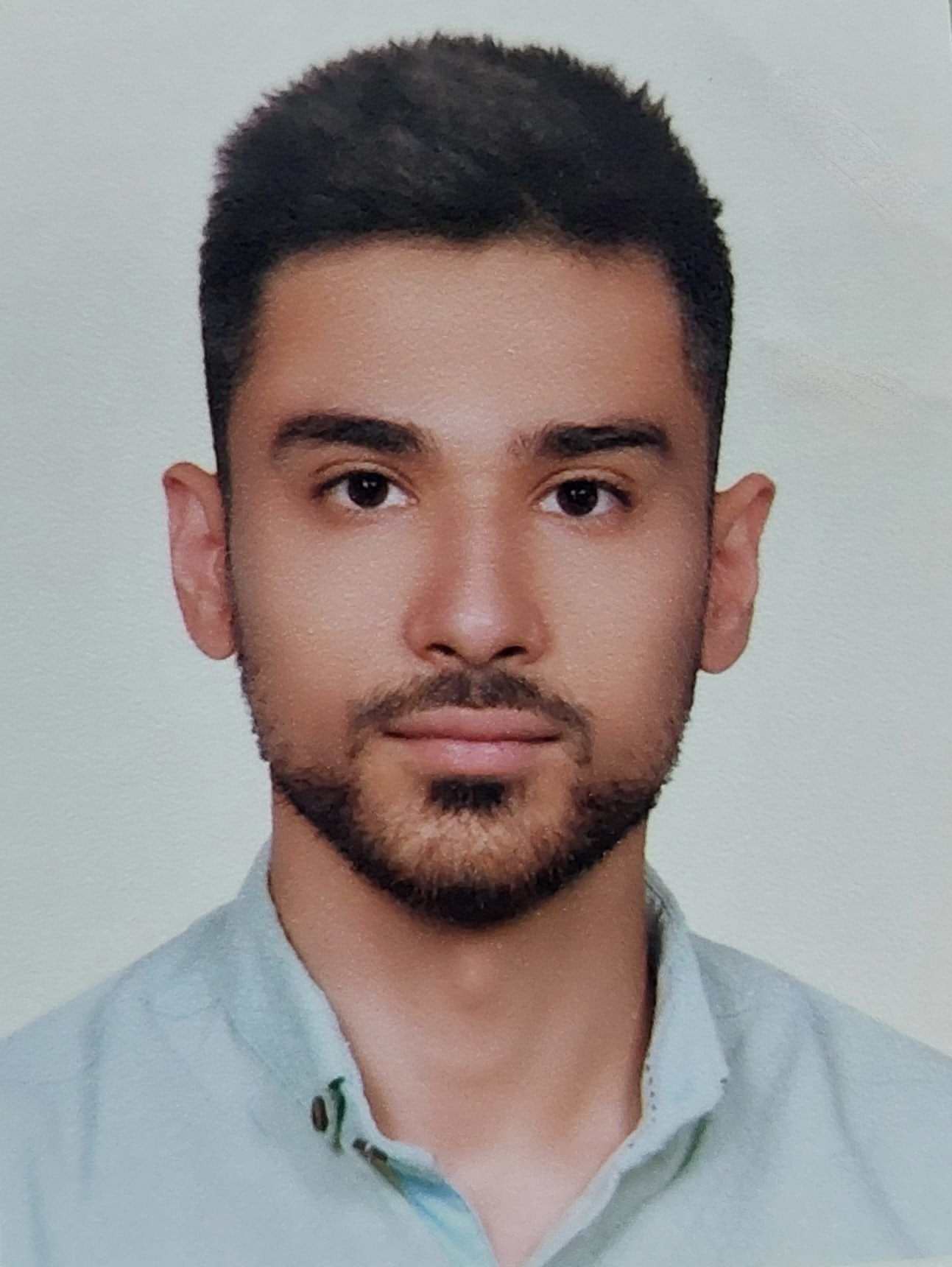}}]{Farshad Zeinali}
(Student Member, IEEE) is a Ph.D. candidate at the Institute for Communication Systems, University of Surrey, United Kingdom. He received his B.Sc. in Electrical Engineering from Golestan University, Gorgan, Iran, in 2019, and his M.Sc. in Telecommunication Engineering from Tarbiat Modares University, Tehran, Iran, in 2023. He worked as a researcher at the Pasargad Institute for Advanced Innovative Solutions, Tehran, Iran, in 2024. His research interests include applying AI/RL algorithms to enhance wireless communication networks and network optimization, particularly resource allocation.
\end{IEEEbiography}
%\vspace{-1.2cm}
\begin{IEEEbiography}[{\includegraphics[width=1in,height=1.25in,clip,keepaspectratio]{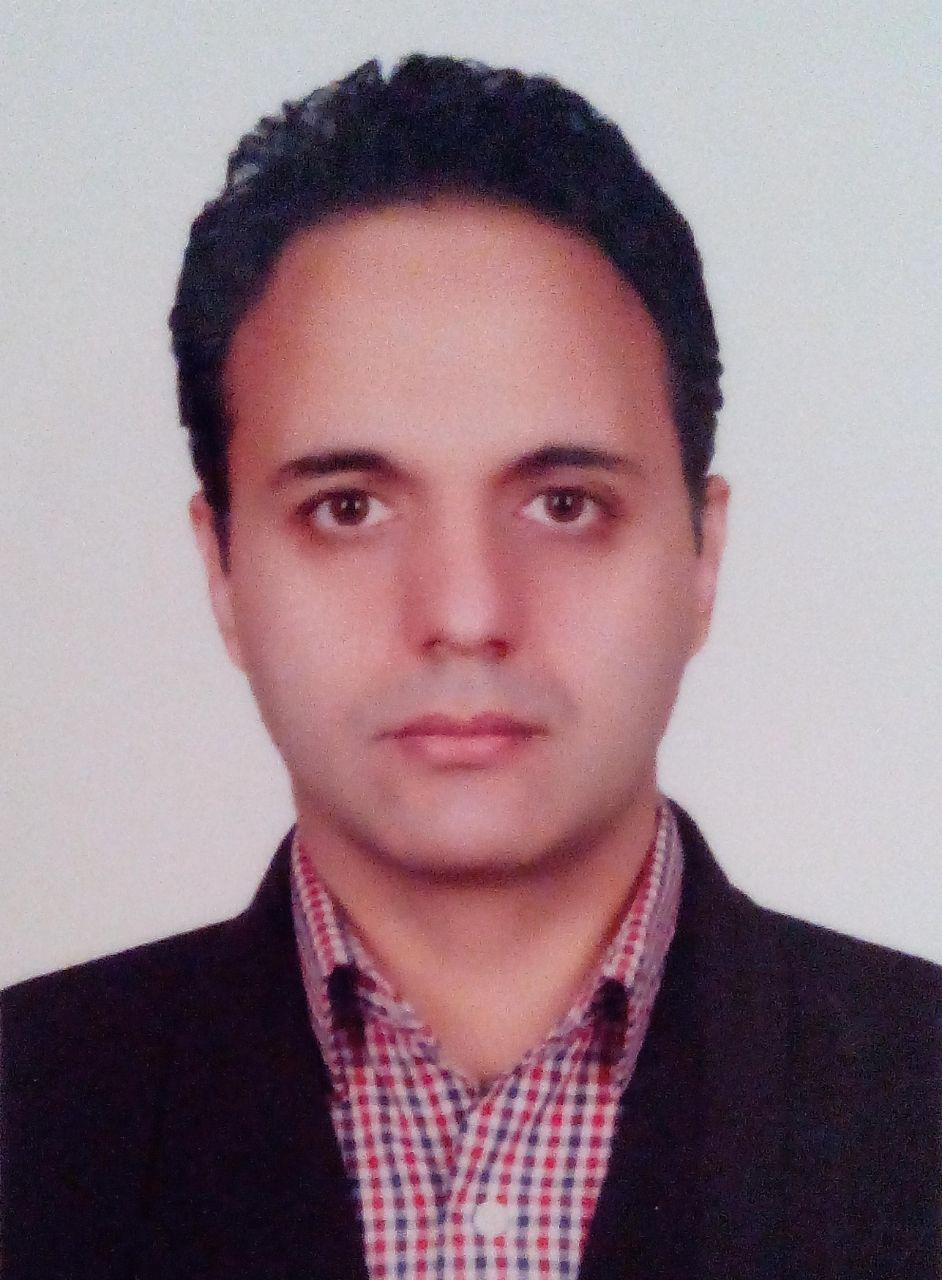}}]{Mohammad Robat Mili}
received the Ph.D. degree in electrical and electronic engineering from the University of Manchester, U.K., in 2012. He held postdoctoral research positions at the Department of Telecommunications and Information Processing, Ghent University, Belgium and the Department of Electrical Engineering, Sharif University of Technology, Iran. His main research interests are in the area of design and analysis of wireless communication networks with particular focus on 5G and 6G cellular networks using mathematical methods such as optimization theory, game theory, and machine learning
\end{IEEEbiography}
%\vspace{-1.2cm}
\begin{IEEEbiography}[{\includegraphics[width=1in,height=1.25in,clip,keepaspectratio]{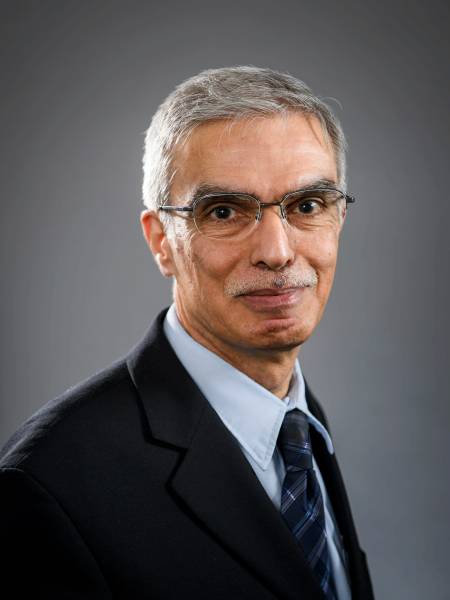}}]{Mohammad Ghavami}
(Senior Member, IEEE) is currently a Professor in telecommunications at London South Bank University. Prior to this appointment, he was with King’s College London from 2002 to 2010 and the Sony Computer Science Laboratories in Tokyo from 2000 to 2002. He has authored the books \emph{Ultra Wideband Signals and Systems in Communication Engineering} and \emph{Adaptive Antenna Systems}. He has also published over 180 technical papers mainly related to UWB and its medical applications. He holds three U.S. and one European patents. He won the esteemed European Information Society Technologies Prize in 2005 and two invention awards from Sony. He has been the Guest Editor of the \emph{IET Proceedings Communications} Special Issue on Ultra Wideband Systems and the Associate Editor of the Special Issue of the \emph{IEICE Journal on UWB Communications}.
\end{IEEEbiography}

\end{document}